\documentclass[manuscript,twocolumn]{aastex701}

\usepackage{amsmath,bm}
\usepackage{placeins}
\usepackage{booktabs}
\hypersetup{hidelinks}

\newcommand\ba{\begin{eqnarray}}
\newcommand\ea{\end{eqnarray}}

\shorttitle{Wavelet-Scattering Signatures of Fuzzy Dark Matter}
\shortauthors{Shimabukuro et al.}
\submitjournal{ApJ}
\graphicspath{{./}{figures/}}

\begin{document}

\title{Wavelet-Scattering Signatures of Fuzzy Dark Matter in Simulated 21\,cm Brightness-Temperature Maps}

\correspondingauthor{Hayato Shimabukuro}

\author{Hayato Shimabukuro}
\affiliation{South-Western Institute for Astronomy Research (SWIFAR), Yunnan University, Kunming, Yunnan 650500, People's Republic of China}
\affiliation{Key Laboratory of Survey Science of Yunnan Province, Yunnan University, Kunming, Yunnan 650500, People's Republic of China}
\affiliation{Graduate School of Science, Division of Particle and Astrophysical Science, Nagoya University, Chikusa-ku, Nagoya 464-8602, Japan}
\email[show]{shimabukuro@ynu.edu.cn (Hayato Shimabukuro)}

\author{Shihang Liu}
\affiliation{Guangxi Key Laboratory for Relativistic Astrophysics, School of Physical Science and Technology, Guangxi University, Nanning 530004, People's Republic of China}
\affiliation{School of Physics and Astronomy, Beijing Normal University, Beijing 100875, People's Republic of China}
\email{sehighs@163.com}

\author{Bohua Li}
\affiliation{Guangxi Key Laboratory for Relativistic Astrophysics, School of Physical Science and Technology, Guangxi University, Nanning 530004, People's Republic of China}
\email[show]{\\bohuali@gxu.edu.cn (Bohua Li)}

\begin{abstract}
\sloppy
We study the effect of fuzzy dark matter (FDM) on the multiscale morphology of simulated redshifted 21\,cm maps during Cosmic Dawn and the Epoch of Reionization. Using FDM-modified \texttt{21cmFAST} light cones, we apply the two-dimensional wavelet scattering transform (WST) to matched 2\,MHz map products. The first-order coefficients $S_1(j)$ summarize wavelet-band amplitudes, while the normalized second-order ratio $R=S_2/S_1(j_1)$ measures ordered cross-scale modulation. FDM shifts and reshapes both summaries through delayed halo and source formation. We compare a two-dimensional power spectrum (PS), WST, and their combination on the same transferred and noisy maps. In this controlled local Fisher analysis, PS+WST reduces the marginalized errors on the FDM mass, effective X-ray emissivity normalization, and ionizing efficiency by approximately a factor of 1.8 relative to the matched two-dimensional PS baseline, although the dominant mass-heating degeneracy remains. An idealized three-wedge test shows that $R$ is less reshaped at the coefficient level than $S_1$. The covariance includes thermal noise conditional on one fiducial light cone but not cosmic variance or foreground residuals; the results are therefore relative information comparisons, not survey forecasts or a demonstration of superiority over a full three-dimensional PS analysis.
\end{abstract}


\section{Introduction}
\label{sec:introduction}

The nature of dark matter remains one of the central open questions in cosmology. While the cold dark matter (CDM) paradigm successfully explains large-scale structure, it faces well-known challenges on sub-galactic scales, such as the cusp-core problem and the apparent deficit of observed dwarf galaxies relative to CDM predictions \citep[e.g.][]{2015PNAS..11212249W,2017ARA&A..55..343B}. Fuzzy dark matter (FDM), or scalar field dark matter, has emerged as a theoretically motivated alternative that may alleviate some of these small-scale tensions \citep[e.g.][]{2000ApJ...534L.127P,2000PhRvL..85.1158H,2002PhLB..545...17B,2014PhRvD..89h3536L,2017PhRvD..95d3541H,2016PhR...643....1M,2025SCPMA..6880409Y,2025FrASS..1238434C,2026PhRvD.114b3522C}. In this scenario, dark matter consists of ultralight bosons with masses around {$m_{\rm FDM} \sim 10^{-22}\,\mathrm{eV}$. Their de~Broglie wavelength can be comparable to galactic scales, so the dark matter behaves more like a coherent wave than as an ensemble of classical particles. Such fields arise naturally in high-energy theories \citep[e.g.][]{2010PhRvD..81l3530A,2016PhR...643....1M} and provide a minimal extension to $\Lambda$CDM without introducing exotic dark-sector interactions.

Key features of FDM include a strong suppression of small-scale structures on both linear and nonlinear scales, together with the formation of solitonic cores in collapsed halos. Linear analyses show that beyond a critical wavenumber set by the particle's de~Broglie, or Jeans, scale, density fluctuations are stabilized by quantum pressure, producing a sharp cutoff in the matter power spectrum \citep{2000PhRvL..85.1158H,2016PhR...643....1M}. On nonlinear scales, both analytical studies of the Schrödinger-Poisson system \citep{2022PhRvD.106j3532T} and high-resolution simulations show that halos below the Jeans mass are suppressed, while those that do form can develop stable soliton-like cores rather than the steep cusps found in CDM \citep{2014NatPh..10..496S,2014PhRvL.113z1302S,2015PhRvD..91l3520M,2020arXiv200704119M,2021MNRAS.506.2603M,2023MNRAS.524.4256M,2025PhRvL.135f1002L,2026LRCA...12....1S}. Recent morphology studies further show that this small-scale suppression reshapes the cosmic web itself \citep{2023MNRAS.525..348D,2025A&A...696A.145Z}. Taken together, halo suppression and core flattening may help address several long-standing small-scale tensions of CDM, including the missing-satellite, cusp-core, and too-big-to-fail problems, while leaving the large-scale successes of CDM largely intact.

Current observational constraints are strong but model dependent. Lyman-$\alpha$ forest analyses give 95\% lower limits on the boson mass, $m_{\rm FDM}$, ranging from $2.0\times10^{-21}$ to $2.0\times10^{-20}\,{\rm eV}$ under different thermal-history and intergalactic-medium assumptions \citep{2017PhRvL.119c1302I,2021PhRvL.126g1302R}, while Milky Way satellite counts give $m_{\rm FDM}>2.9\times10^{-21}\,{\rm eV}$ for the adopted galaxy-halo and selection model \citep{2021PhRvL.126i1101N}. These bounds should not be combined without accounting for their different assumptions. In this work, we use $m_{\rm FDM}=10^{-21}$\,eV and $5\times10^{-21}$\,eV as controlled morphology benchmarks rather than preferred values.

The redshifted 21\,cm line of neutral hydrogen traces the thermal and ionization history of the intergalactic medium from Cosmic Dawn through reionization \citep[e.g.,][]{2012RPPh...75h6901P,2002ApJ...579....1F,2012MNRAS.422..926M,2021ApJ...918...14L,2023PASJ...75S...1S}. Because the formation of the first stars and galaxies depends on the abundance of low-mass halos, the 21\,cm signal provides a direct probe of how FDM suppresses early structure formation. In FDM cosmologies, the deficit of small halos delays the emergence of the first luminous sources and shifts the characteristic 21\,cm absorption and emission features to lower redshift \citep[e.g.,][]{2014JCAP...06..011K,Nebrin2019,2021ApJ...913....7J,2022PhRvD.106d3529H,2022PhRvD.106f3504F,2025PhRvD.112j3534L}. The 21\,cm power spectrum therefore provides the most direct conventional reference for the present work: it exhibits reduced small-scale fluctuation power and redshift-shifted features in FDM models, allowing observations at $10\lesssim z\lesssim30$ to probe wave-like dark matter through its effect on early structure formation at epochs inaccessible to many low-redshift tracers.

Complementary to power-spectrum analyses, the 21\,cm forest, consisting of narrow absorption lines in the spectra of high-redshift radio sources, offers a complementary probe of dark matter. Each line traces intervening neutral hydrogen in small structures such as filaments or mini-halos, directly sampling the small-scale matter distribution during Cosmic Dawn. Because FDM suppresses fluctuations below the de~Broglie scale, it reduces the number and depth of detectable absorption lines relative to CDM predictions. Recent studies show that 21\,cm forest statistics can constrain the FDM particle mass or distinguish it from other dark matter models \citep[e.g.,][]{2014PhRvD..90h3003S,2020PhRvD.101d3516S,2020PhRvD.102b3522S,2023PhRvD.107l3520S,2026PhRvD.113h3525S}. More recently, the wavelet scattering transform (WST) has been applied to 21\,cm fields as a multiscale summary beyond the power spectrum (PS) \citep[e.g.][]{2022MNRAS.513.1719G,2023MNRAS.519.5288G,2024A&A...686A.212H,2025PhRvD.112f3557S}. The first-order statistic $S_1$ summarizes wavelet-band amplitudes, while $S_2$ measures modulation of one wavelet response by a broader scale. We use $R=S_2/S_1(j_1)$ as a reduced ordered statistic; it is sensitive to higher-order morphology but is not phase-only.

The present work is designed as a morphological extension of existing FDM 21\,cm calculations. The reference power-spectrum trends used below reproduce the physical delay caused by the same FDM-modified \texttt{21cmFAST} pipeline, while the new element is a WST-based characterization of how this delay and small-scale suppression appear in two-dimensional brightness-temperature maps. In contrast to a purely descriptive WST study, we explicitly compare the WST summaries with the power spectrum and test whether the two summaries carry complementary Fisher information.

We first characterize the redshift evolution of $S_1$ and $R$ in CDM and FDM maps. We then compare matched two-dimensional PS, WST, and combined data vectors using the same 2\,MHz maps, instrumental transfer function, integration time, and thermal-noise realizations. Finally, we test three idealized foreground wedges and an alternative second-order normalization. WST receives no additional observing time, although real image-domain inference would require stronger calibration and mapmaking control than the present linear-map experiment.

The calculation uses a minimal source model and an analytic SKA1-Low-inspired instrumental-transfer model rather than a physical array layout. Foreground residuals, calibration errors, chromatic beams, polarization leakage, and cosmic variance are not modeled. The supported result is therefore complementarity relative to a matched two-dimensional PS baseline, not a survey reach or a comparison with a full three-dimensional PS analysis.

The paper is organized as follows.
Section~\ref{sec:FDM} reviews the theoretical framework of FDM and summarizes the modifications to the linear power spectrum and halo mass function (HMF) implemented in our simulations.
Section~\ref{sec:21cm} describes the basic 21\,cm physics and the simulation of 21\,cm brightness-temperature cubes using the modified \texttt{21cmFAST} code.
Section~\ref{sec:wst} introduces the wavelet scattering transform formalism and outlines its numerical implementation.
Section~\ref{sec:results} presents the intrinsic WST responses, the controlled instrumental-transfer and noise tests, and the foreground-wedge response. Section~\ref{sec:fisher} gives the matched Fisher comparison, and Section~\ref{sec:discussion} summarizes the results. Appendices~\ref{app:fdm_background} through \ref{app:alpha_robustness} provide theoretical and simulation diagnostics, while Appendices~\ref{app:normalization} and \ref{app:validation} give the normalization and Fisher-validation tests.

\section{Fuzzy Dark Matter Framework}
\label{sec:FDM}

FDM is modeled as an ultralight real scalar field with mass
\(m_{\rm FDM}\sim 10^{-22}\,{\rm eV}\), whose de~Broglie wavelength can reach
kiloparsec scales. At such small masses the wave character of the field becomes
manifest on astrophysical scales through interference, so that structure
formation deviates from purely collisionless CDM on small
scales \citep{2000PhRvL..85.1158H}. On large scales FDM follows CDM, whereas on
subgalactic scales the gradient (``quantum pressure'') term generates an
effective Jeans scale below which gravitational collapse is suppressed.

For the purposes of the 21\,cm simulations, the essential FDM ingredients are the linear small-scale cutoff and the resulting suppression of the HMF. We therefore keep the field-theory background only in compact form here and focus on the quantities that enter the modified \texttt{21cmFAST} pipeline. A self-contained derivation of the equation of motion for the scalar field, its nonrelativistic limit, 
and the hydrodynamic interpretation of the quantum-pressure term is given in Appendix~\ref{app:fdm_background}.

\paragraph{Linear regime and power-spectrum cutoff.}

For a Fourier mode with comoving wavenumber $k$, linearizing the
Schr\"odinger-Poisson system gives \citep{2017PhRvD..95d3541H}
\begin{equation}
\ddot\delta_k+2H\dot\delta_k+
\left(\frac{\hbar^2 k^4}{4m_{\rm FDM}^2a^4}-4\pi G\bar\rho\right)\delta_k=0.
\label{eq:fdm_linear_growth}
\end{equation}
Here $\delta_k$ is the Fourier amplitude of the matter-density contrast at comoving wavenumber $k$, $H=\dot a/a$ is the Hubble parameter, $a$ is the scale factor, $m_{\rm FDM}$ is the FDM particle mass, $\bar\rho$ is the homogeneous matter density, $G$ is Newton's constant, and $\hbar$ is the reduced Planck constant.
The corresponding comoving and physical Jeans wavenumbers are
\begin{equation}
k_J(a)=a\left(\frac{16\pi G\bar\rho m_{\rm FDM}^2}{\hbar^2}\right)^{1/4},
\qquad k_{J,{\rm phys}}=\frac{k_J}{a}.
\end{equation}
Here $k_J(a)$ and $k_{J,{\rm phys}}$ denote the comoving and physical Jeans wavenumbers, respectively. Thus $\lambda_{J,{\rm phys}}=2\pi/k_{J,{\rm phys}}$ is the physical Jeans wavelength and $m_J=(4\pi/3)\bar\rho[\lambda_{J,{\rm phys}}/2]^3$ is the corresponding characteristic mass. Modes below this
physical length are stabilized by the gradient term.

The corresponding FDM linear-matter transfer-function fit of \citet{2000PhRvL..85.1158H} is
\begin{equation}
\frac{P_{\mathrm{FDM}}(k)}{P_{\mathrm{CDM}}(k)}
 = \left[\frac{\cos\!\big(x^3\big)}{1+x^8}\right]^2,
\qquad
x = 1.61\,m_{22}^{1/18}\!\left(\frac{k}{k_{\mathrm J,eq}}\right).
\label{eq:fdm_transfer_redux}
\end{equation}
Here $P_{\rm FDM}(k)$ and $P_{\rm CDM}(k)$ are the linear matter power spectra in the FDM and CDM models, respectively, $x$ is the dimensionless fitting variable shown above, $m_{22}\equiv m_{\rm FDM}/(10^{-22}\,{\rm eV})$, and $k_{J,{\rm eq}}\simeq9\,m_{22}^{1/2}\,{\rm Mpc}^{-1}$ is the comoving Jeans wavenumber at matter-radiation equality. Both $k$ and $k_{J,{\rm eq}}$ are comoving wavenumbers. The FDM linear-matter transfer function and the HMF prescription below are incorporated in the 21~cm signal simulation. The WST does not measure the linear cutoff directly, but responds indirectly through the delayed and spatially altered source fields.
The corresponding characteristic half-mode wavenumber for this fit is
\(k_{1/2}\simeq4.5\,m_{22}^{4/9}\,{\rm Mpc}^{-1}\). The suppression is sharper
than the commonly adopted warm-dark-matter transfer shape. In
nonlinear FDM simulations, the associated wave dynamics can suppress the
abundance of low-mass halos and produce solitonic central profiles.

\paragraph{Nonlinear Structure Formation in Fuzzy Dark Matter.}

In the nonlinear regime, the wave dynamics of FDM suppress
the formation and abundance of low-mass halos and can generate
solitonic cores in virialized objects. The semi-numerical
treatment used in this work does not resolve internal halo
profiles. Nonlinear FDM effects enter the
\texttt{21cmFAST} calculation through the simulation-motivated
global HMF and its environmental modulation. Whether FDM
alleviates observed dwarf-galaxy abundance or central-density
tensions depends additionally on the particle mass, halo
evolution, observational selection, and baryonic physics.

The suppression of the low-mass FDM HMF reflects both the
cutoff in the linear initial power spectrum and the subsequent
nonlinear wave dynamics. These contributions should be
distinguished. Simulations evolved from the FDM initial power
spectrum without the full wave dynamics produce an intermediate
suppression, whereas the full FDM treatment gives a stronger
reduction in the abundance of low-mass halos
\citep{2021MNRAS.506.2603M,2023MNRAS.524.4256M,
2025PhRvD.112j3534L}.

We adopt the simulation-motivated fitting formula for the global FDM HMF,
\begin{equation}
\left.\frac{{\mathrm d}n}{{\mathrm d}m}\right|_{\mathrm{FDM}}
=
\left.\frac{{\mathrm d}n}{{\mathrm d}m}\right|_{\mathrm{CDM}}
\left[
1+
\left(\frac{m}{m_0}\right)^{\alpha}
\right]^{-2.2},
\label{eq:fdm_hmf}
\end{equation}
where $m$ denotes halo mass, $m_0=1.6\times10^{10}\,m_{22}^{-4/3}\,M_{\odot}$ is the characteristic halo-mass scale of the fitting formula, and the fiducial suppression-shape index is $\alpha=-1.1$ \citep{2016ApJ...818...89S}. The quantity $\mathrm{d}n/\mathrm{d}m$ is the differential halo mass function.
Here the CDM factor
$\left.(\mathrm{d}n/\mathrm{d}m)\right|_{\mathrm{CDM}}$
is evaluated using the standard excursion-set construction and
the unsuppressed CDM variance
$\sigma_{\mathrm{CDM}}$
\citep{press1974formation,bond1991excursion,
sheth1999large,sheth2001ellipsoidal}.
Eq.~(\ref{eq:fdm_hmf}) defines the global,
unconditional FDM HMF averaged over the simulation volume
\citep{2021MNRAS.506.2603M,2023MNRAS.524.4256M}.

Analytic excursion-set prescriptions using suppressed
linear power spectra provide alternative estimates of the FDM
HMF. However, such estimates can predict a substantially
different low-mass turnover from the available full-wave
simulations. We therefore retain the simulation-motivated fitting
form adopted in the FDM-modified
\texttt{21cmFAST} pipeline \citep{2025PhRvD.112j3534L}.
This fitting form is broadly consistent with 
the state-of-the-art cosmological simulations
of the Schr\"odinger-Poisson system, 
although its quantitative accuracy and redshift dependence remain limited by
the available simulation volumes and halo statistics 
\citep{2023MNRAS.524.4256M}.

The index $\alpha$ is not a fundamental particle parameter
like $m_{22}$. It controls the shape of the phenomenological
low-mass HMF suppression. We fix $\alpha=-1.1$ in the primary
local Fisher comparison and show the descriptive WST response
to variations of $\alpha$ in
Appendix~\ref{app:alpha_robustness}.

To go beyond the global HMF
and model spatial variations in halo abundance, we adopt the
environmental-modulation ansatz introduced by \citet{2025PhRvD.112j3534L}.
Particularly, the peak-height variable that enters the CDM HMF factor
in Eq.~(\ref{eq:fdm_hmf}) is replaced by
\begin{equation}
\nu^2
=
\frac{
[\delta_c-\delta_{\rm FDM}(z)]^2
}{
\sigma^2_{\rm CDM}(m,z)
-
\sigma^2_{\rm FDM}(M,z)
}.
\label{eq:fdm_env}
\end{equation}
Here $\nu$ is the excursion-set peak-height variable, $m$ is the halo mass, $M$ is the total mass of the excursion-set region, $\delta_c$ is the critical linear overdensity for collapse, $\delta_{\rm FDM}(z)$ is the linear FDM overdensity smoothed over that region, and $\sigma^2_{\rm CDM}(m,z)$ and $\sigma^2_{\rm FDM}(M,z)$ are the corresponding linear-theory density variances on the halo and region mass scales.

Eq.~(\ref{eq:fdm_env}) does not describe a separate HMF fit. 
It introduces the environment-dependent response of the
global FDM HMF through its CDM HMF factor and is used to
construct the local collapsed fraction and source field. In the
large-region limit, $M\rightarrow\infty$,
the regional overdensity and smoothed variance vanish, and the
density-modulated HMF reduces to the global HMF in
Eq.~(\ref{eq:fdm_hmf}).

The FDM linear-matter transfer function in
Eq.~(\ref{eq:fdm_transfer_redux}), the global HMF in
Eq.~(\ref{eq:fdm_hmf}), and the environmental modulation in
Eq.~(\ref{eq:fdm_env}) together determine the density field and
the local collapsed fractions used to construct the radiative
source fields in the modified \texttt{21cmFAST} calculation.
These quantities form the basis for the subsequent simulation
of the 21\,cm brightness-temperature field during Cosmic Dawn
and the Epoch of Reionization (EoR).

\section{The 21 cm Signal as a Probe of Cosmic Structure Formation}
\label{sec:21cm}
The redshifted 21\,cm line of neutral hydrogen serves as a unique tracer of the intergalactic medium throughout cosmic history, from the Dark Ages to the completion of reionization \citep[e.g.][]{fur,2012RPPh...75h6901P,2013PhRvD..87d3002L,2023PASJ...75S...1S}. The observable quantity is the differential brightness temperature of the 21\,cm transition against the cosmic microwave background (CMB), expressed as
\begin{align}
\delta T_{\mathrm b}(\nu)
 &\simeq 27\,x_{\mathrm{HI}}\,(1+\delta_{\mathrm b})
 \!\left(\frac{\Omega_{\mathrm b}h^2}{0.023}\right)
 \!\left(\frac{0.15}{\Omega_{\mathrm m}h^2}
 \frac{1+z}{10}\right)^{1/2}  \notag\\[4pt]
 &\times
 \!\left(1-\frac{T_\gamma}{T_{\mathrm S}}\right)
 \mathrm{mK},
\label{eq:dTb_basic}
\end{align}
where $x_{\mathrm{HI}}$ is the neutral hydrogen fraction, $\delta_{\mathrm b}$ the baryon overdensity, $T_{\mathrm S}$ the spin temperature of the 21\,cm transition, and $T_\gamma$ the CMB temperature at redshift $z$. Here $\nu$ is the observed frequency, $\Omega_{\rm b}$ and $\Omega_{\rm m}$ are the present-day baryon and total-matter density parameters, respectively, and $h$ is the dimensionless Hubble constant defined by $H_0=100h\,{\rm km\,s^{-1}\,Mpc^{-1}}$.
The sign of $\delta T_{\mathrm b}$ depends on the ratio between $T_{\mathrm S}$ and $T_\gamma$:
the signal appears in absorption when $T_{\mathrm S}<T_\gamma$,
and in emission when $T_{\mathrm S}>T_\gamma$.
Spatial variations of these quantities reflect the progress of structure formation
and the radiative processes that accompany it.

The evolution of the 21\,cm signal is governed mainly by three astrophysical processes. First, ultraviolet photons from the earliest stars couple the spin temperature to the kinetic temperature of the neutral gas through resonant Ly$\alpha$ scattering, the Wouthuysen-Field effect \citep{1952AJ.....57...31W,1958PIRE...46..240F,1959ApJ...129..536F}.
This marks the Ly$\alpha$ coupling epoch and the beginning of the so-called Cosmic Dawn.
As more energetic sources form, soft X-ray photons emitted by stellar remnants
and early accreting black holes gradually heat the intergalactic medium.
This X-ray heating epoch raises the kinetic temperature above the CMB temperature
and transforms the 21\,cm signal from absorption to emission.
Eventually, ultraviolet radiation from galaxies ionizes the surrounding gas,
reducing the neutral fraction $x_{\mathrm{HI}}$ and terminating the 21\,cm signal.
The morphology of ionized bubbles and the evolution of the mean neutral fraction
encode the large-scale topology of reionization.

Fluctuations of $\delta T_{\mathrm b}$ arise from spatial variations
in density, velocity, and radiation backgrounds.
They are often characterized by the dimensionless 21\,cm power spectrum,
\begin{equation}
\Delta^2_{21}(k,z)\equiv \frac{k^3 P_{21}(k,z)}{2\pi^2}.
\label{eq:p21}
\end{equation}
Here $P_{21}(k,z)$ is the three-dimensional 21\,cm brightness-temperature power spectrum and $k$ is the comoving wavenumber. This quantity measures the variance of brightness-temperature fluctuations
per logarithmic interval in wavenumber.
Different astrophysical epochs produce distinct peaks in $\Delta^2_{21}(k,z)$:
the Ly$\alpha$ coupling peak, the X-ray heating peak, and the reionization peak.
The 21\,cm field spans a broad range of scales, although the accessible modes depend strongly on the instrument and foreground treatment. FDM first changes the linear matter spectrum and HMF; the resulting source delay then changes the timing and amplitude of $\Delta^2_{21}(k,z)$. Interferometers such as SKA-Low and HERA may therefore constrain such source-mediated effects, but the controlled mock below does not establish a boson-mass reach.

Power-spectrum measurements are the closest conventional counterpart to the WST analysis developed in this paper, because both statistics are extracted from spatial fluctuations of the brightness-temperature field. Future tomographic imaging of the 21\,cm signal will further enable direct mapping of ionized, heated, and Ly$\alpha$-coupled structures, allowing their morphology to be compared with theoretical predictions based on FDM and other alternative dark matter models.

To model the evolution of the 21\,cm signal, we employ the publicly available semi-numerical code \texttt{21cmFAST} \citep{2011MNRAS.411..955M,2019MNRAS.484..933P},
which self-consistently follows the formation of the first luminous sources
and the subsequent reionization and heating of the intergalactic medium.

We adopt this phenomenological \texttt{21cmFAST} parameterization
because the FDM-modified pipeline used here, 
including the modified power spectrum, HMF suppression, and environmental modulation, was constructed and validated in this framework. This choice keeps the WST analysis directly comparable to the underlying FDM 21\,cm simulations and makes the Fisher calculation computationally tractable. More recent source models in newer versions of \texttt{21cmFAST} include additional astrophysical parameters and feedback prescriptions, but incorporating the same FDM HMF prescription into those models would require a separate implementation and validation.

The code computes the brightness temperature field $\delta T_{\mathrm b}(\mathbf{x},z)$
on large cosmological volumes by combining density fields generated from
Gaussian initial conditions with approximate prescriptions
for astrophysical radiation and feedback processes.

The reionization process is described by the photon-counting criterion
\begin{equation}
\zeta\, f_{\mathrm{coll}}(\mathbf{x},z;R,m_{\min})\ge1,
\label{eq:ionization_condition}
\end{equation}
where $\zeta$ is the ionizing efficiency and
$f_{\mathrm{coll}}$ represents the collapsed fraction of matter in halos
above a minimum mass $m_{\min}$ within a comoving region of radius $R$.
Here $\mathbf{x}$ denotes comoving position, $z$ is redshift, $R$ is the excursion-set filtering radius, and $m_{\min}$ is the minimum halo mass included in the collapsed fraction. A region is considered ionized when this condition is satisfied.
The parameters $\zeta$, $T_{\mathrm{vir}}$, and $L_{\mathrm X}/\mathrm{SFR}$ describe the effective ionizing efficiency, the minimum halo mass for star formation, and the specific X-ray luminosity per unit star formation rate.
The source model documented by \citet{2025PhRvD.112j3534L}
uses the following fiducial values:
$\zeta=20$, $T_{\rm vir}=2\times10^4\,{\rm K}$, and $L_\mathrm{X}/{\rm SFR}
=10^{40}\,{\rm erg}\,\mathrm{s}^{-1}\,\mathrm{keV}^{-1}\,(M_\odot\,{\rm yr}^{-1})^{-1}$.

\citet{2025PhRvD.112j3534L} modified this framework to incorporate
both the linear and nonlinear effects of FDM on structure formation.
Specifically, the authors implemented the FDM linear power spectrum
and the corresponding HMF derived from full numerical simulations,
together with an environmental modulation prescription
that accounts for density-dependent variations in halo abundance.
These improvements enabled predictions of the 21\,cm signal and its power spectrum for a wide range of FDM particle masses, conventionally expressed through $m_{22}$.

\subsection{Light-Cone Construction and Source-Model Assumptions}
\label{subsec:lightcones}

Each simulated light cone analyzed in this work consists of $200\times200\times2962$ cells: $200\times200$ transverse cells and 2962 native cells along the line of sight. The transverse extent is 300 comoving Mpc, the light cone spans $4.00\leq z\leq35.05$, and the transverse pixel size is 1.5 comoving Mpc. For the Fisher analysis, we average
the native line-of-sight cells into nonoverlapping 2\,MHz slabs.
Over the primary analysis range $8\leq z\leq25$, each slab contains
23 to 42 native cells, corresponding to a line-of-sight depth of
approximately 34 to 63 comoving Mpc. This procedure yields 17
nonoverlapping slab centers. Three additional slabs at $z>25$ are
used only to examine the sensitivity of the results to the
highest-redshift portion of the light cone.

We adopt the simulations performed in \citet{2025PhRvD.112j3534L}
that use the legacy mass-independent source prescription. 
In this prescription, the ionizing efficiency, $\zeta$, combines the effects
of star formation efficiency, the ionizing photons produced per stellar baryon, 
and the escape fraction of ionizing photons. 
In other words, the star formation efficiency, $f_\star$, is absorbed
in $\zeta$ when high-$z$ galaxies as the ionizing sources are concerned.

Meanwhile, the effects from the X-ray sources are described by 
the specific X-ray luminosity per unit star formation escaping host galaxies,
$L_\mathrm{X}/{\rm SFR}$.
It is analogous to the product of 
the number of photons produced per stellar baryon and the escape fraction,
and one should additionally specify the star formation efficiency;
see Eq.~(5) in \citet{2018MNRAS.477.3217G}.
In \texttt{21cmFAST}, the default value of $f_\star\approx0.05$ is held fixed
in all our simulations.

The source model adopted in this work
includes star formation only in atomic-cooling halos,
with the minimum source-hosting halo mass controlled by $T_{\rm vir}$.
It does not include molecular-cooling minihalos,
a distinct Population III source population, nor Lyman-Werner feedback. 
These omitted components are expected to
become increasingly important at the highest redshifts.
Therefore, in what follows, we restrict the primary Fisher comparison
to $8\leq z\leq25$. 
The three slabs at $z>25$ are used only as a
limited robustness test and are not interpreted as a complete
description of 21\,cm fluctuations driven by Pop III stars.

The light cones produced with the above model assumptions
provide the 21\,cm brightness-temperature fields
used throughout the subsequent analysis.
Our simulations also keep track of the redshift evolution of the density,
Ly$\alpha$-coupling, X-ray-heating, and ionization fields
generated by the specified FDM and source models, 
subject to the limitations described above.

\section{Wavelet Scattering Transform}
\label{sec:wst}

We apply the two-dimensional WST to the 21\,cm brightness-temperature fluctuations, $\delta T_b(\mathbf{x},z)$, where $\mathbf{x}$ denotes the sky-plane coordinates at a given redshift $z$. The WST provides a set of translation-invariant, or shift-insensitive, and deformation-stable summary statistics by cascading wavelet convolutions, non-linear modulus operations, and spatial averaging. This formalism captures both Gaussian and non-Gaussian features of the signal in a computationally efficient way.

Given a two-dimensional field $I(\mathbf{x})=\delta T_b(\mathbf{x},z)$, we define the spatial average
\begin{equation}
\langle f \rangle \equiv \frac{1}{A}\int \mathrm{d}^2x \, f(\mathbf{x}),
\end{equation}
where $A\equiv\int \mathrm{d}^2x$ is the map area (or, in the discrete implementation, $\langle f\rangle$ corresponds to an average over pixels). Here $f(\mathbf{x})$ denotes an arbitrary map-domain function and $\mathbf{x}$ is the two-dimensional sky-plane coordinate.
With this convention, the scattering coefficients are insensitive to the map area and can be compared consistently across realizations.

The zeroth-order scattering coefficient is the global mean,
\begin{equation}
S_0 = \langle I \rangle.
\end{equation}

The first-order coefficients are obtained by convolving the field with a family of directional wavelets $\psi_{j,\ell}$, indexed by the wavelet scale $j$ and orientation $\ell$, taking the modulus, and then averaging over positions:
\begin{equation}
S_1(j,\ell) \;=\; \left\langle \bigl|\, I * \psi_{j,\ell} \,\bigr| \right\rangle.
\end{equation}
Here $*$ denotes convolution on the two-dimensional map, $j=0,\ldots,J-1$ labels logarithmically spaced spatial scales, and $\ell=0,\ldots,L-1$ labels wavelet orientations. Smaller $j$ corresponds to finer physical structures, while larger $j$ corresponds to broader structures. These directional coefficients measure the typical fluctuation amplitude of $\delta T_b$ at different spatial scales and orientations and can be viewed as a localized, deformation-stable analogue of the power spectrum. In the analysis below, we report the orientation-averaged quantity
\begin{equation}
S_1(j) \equiv \frac{1}{L}\sum_{\ell=0}^{L-1} S_1(j,\ell).
\end{equation}

To probe higher-order correlations and scale couplings, we compute the second-order scattering coefficients,
\begin{equation}
S_2(j_1,j_2,\ell_1,\ell_2)
= \left\langle \left| \, \left| I * \psi_{j_1,\ell_1} \right| * \psi_{j_2,\ell_2} \, \right| \right\rangle.
\end{equation}
These coefficients quantify how the modulus response characterized by $(j_1,\ell_1)$ is modulated at the broader scale $(j_2,\ell_2)$. They can respond to higher-order organization, but are not pure phase statistics.

For the results presented below, we average these directional second-order coefficients over orientations,
\begin{equation}
S_2(j_1,j_2) \equiv \frac{1}{L^2}\sum_{\ell_1=0}^{L-1}\sum_{\ell_2=0}^{L-1} S_2(j_1,j_2,\ell_1,\ell_2),
\end{equation}
and we restrict attention to $j_1<j_2$, so that a finer reference scale is modulated by a broader scale.

In practice, the averaging is implemented via a low-pass filter (followed by subsampling) in the standard scattering construction, as done in \texttt{Kymatio}; the expressions above represent the corresponding continuum notation. In the numerical data vectors used for the Fisher comparison, the zeroth-order coefficient $S_0$ is kept separate and is not included in the $S_1$ block or in the denominator used to form $R=S_2/S_1$.

To quantify cross-scale modulation, we define the normalized second-order ratio
\begin{equation}
R(j_1,j_2) = \frac{S_2(j_1,j_2)}{S_1(j_1)}.
\end{equation}
We use this form because dividing by the reference response $S_1(j_1)$ reduces its amplitude dependence. It does not remove all first-order information, additive noise, or a scale-dependent instrumental transfer function. The measured wedge test shows partial cancellation for some pairs but substantial heterogeneous responses for others. Thus $R$ is treated as an ordered reduced scattering statistic, not as a pure non-Gaussian observable or a bispectrum equivalent. Appendix~\ref{app:normalization} compares it with a symmetric-denominator alternative.

In practice, we compute $S_1(j)$ and $S_2(j_1,j_2)$ for each redshift slice of the simulated brightness-temperature cubes $\delta T_b(x,y,z)$, treating each slice as a two-dimensional input field $I(x,y)$. The coefficients are averaged over spatial positions and, unless stated otherwise, over all wavelet orientations, yielding a compact set of descriptors indexed only by scale $j$ and by pairs $(j_1,j_2)$.

The numerical transform uses \texttt{Kymatio} version \texttt{0.4.0.dev0} \citep{2018arXiv181211214A}. We use $J=6$, $L=4$ for the descriptive curves and $J=5$, $L=4$, \texttt{max\_order=2} for the Fisher vectors, yielding five $S_1$ coefficients and ten ordered ratios in the latter case. Coefficients are selected from the scattering-order and path metadata, with $S_0$ removed before orientation averaging. The measured passbands of the discrete filters are given in Appendix~\ref{app:morlet}; we do not identify the largest $J=6$ scale with the box fundamental mode.

The resulting $S_1$ and $R$ vectors are used as compact summaries of band amplitudes and ordered cross-scale modulation.

\section{Results}
\label{sec:results}

\subsection{WST Coefficients Without Thermal Noise}

The descriptive $J=6$ curves use transverse planes sampled along the simulated light cone and are distinct from the nonoverlapping 2\,MHz bins.

\subsubsection{First-Order WST Coefficient}

We first examine the intrinsic first-order coefficients $S_1(j)$, which summarize orientation-averaged fluctuation amplitudes in the six descriptive wavelet bands. Figure~\ref{fig:wst_1st} compares CDM with $m_{22}=10$ and 50 at fixed $\alpha=-1.1$. FDM shifts and reshapes the main features toward lower redshift, most strongly for $m_{22}=10$, because the suppressed HMF delays the Ly$\alpha$, heating, and ionization histories. These wavelet bands respond to the source-mediated 21\,cm morphology and do not measure the linear FDM cutoff directly. The dependence on the HMF-shape parameter is shown in Appendix~\ref{app:alpha_robustness}; statistical separation is assessed only in Section~\ref{sec:fisher}.

\begin{figure*}
    \centering
    \includegraphics[width=1.0\hsize]{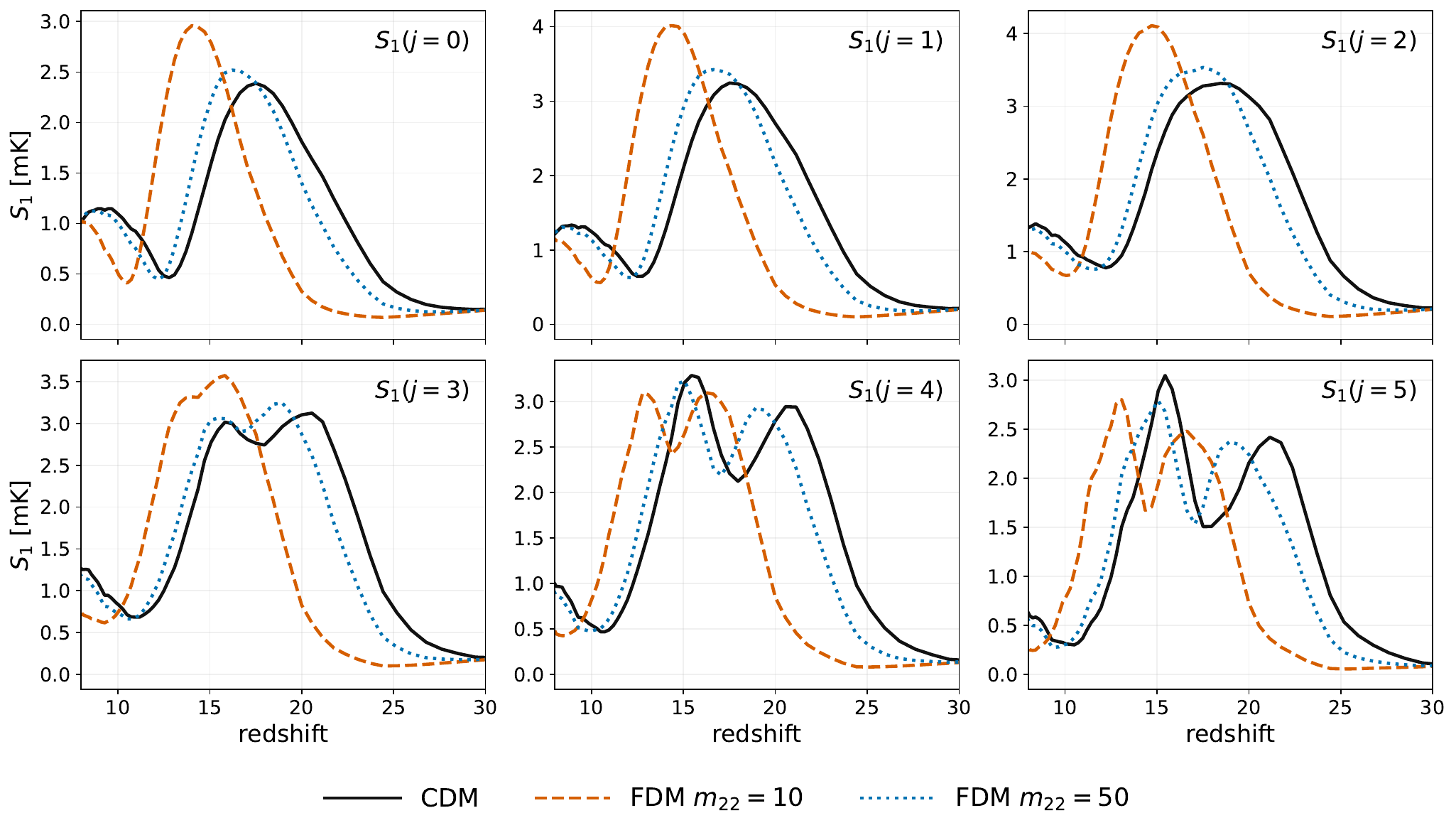}
    \caption{
    First-order WST coefficients without thermal noise.
    Dependence on the FDM mass parameter at fixed $\alpha=-1.1$, comparing CDM with FDM models $m_{22}=10$ and $m_{22}=50$.
    The model-dependent shifts arise indirectly through the HMF and source histories.
    The dependence on the HMF-shape parameter $\alpha$ is shown in Appendix~\ref{app:alpha_robustness}.
    Coefficients are selected by Kymatio path metadata, with $S_0$ excluded.
    }
    \label{fig:wst_1st}
\end{figure*}

Figures~\ref{fig:wst_1st_astro_tvir} through \ref{fig:wst_1st_astro_zeta} show the corresponding astrophysical responses. Increasing $T_{\rm vir}$ delays source formation and shifts the $S_1$ features to lower redshift; increasing $L_\mathrm{X}/{\rm SFR}$ advances and shortens the heating transition; and increasing $\zeta$ primarily suppresses the low-redshift reionization tail. These curves are descriptive sweeps of the adopted source model.

\begin{figure*}
    \centering
    \includegraphics[width=0.9\hsize]{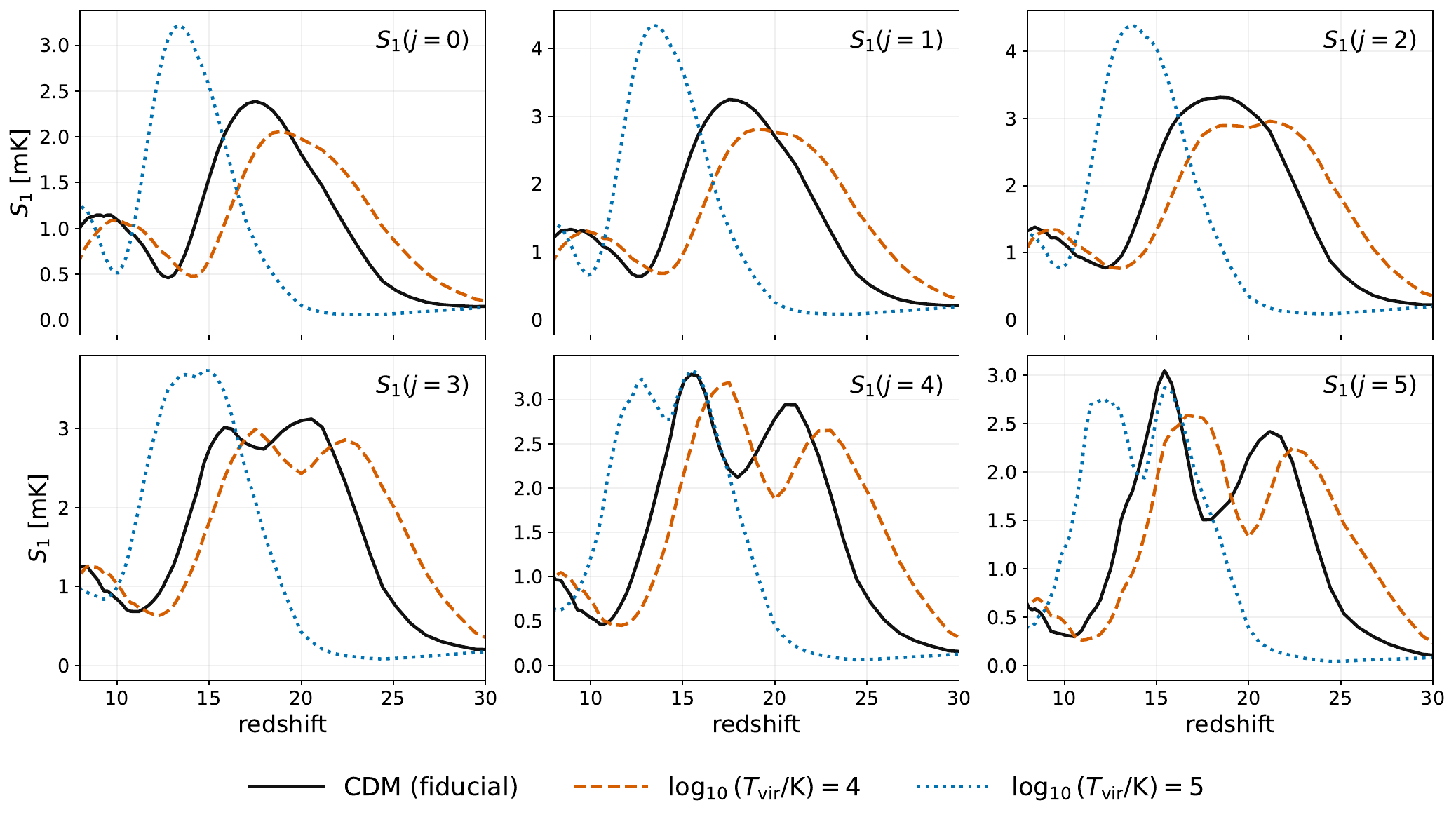}
    \caption{First-order $S_1(j)$ curves from the CDM $T_{\rm vir}$ sweep, labeled $\log_{10}(T_{\mathrm{vir}}/{\rm K})=4$ and 5. This descriptive sweep is not used as the primary Fisher fiducial.}
    \label{fig:wst_1st_astro_tvir}
\end{figure*}

\begin{figure*}
    \centering
    \includegraphics[width=0.9\hsize]{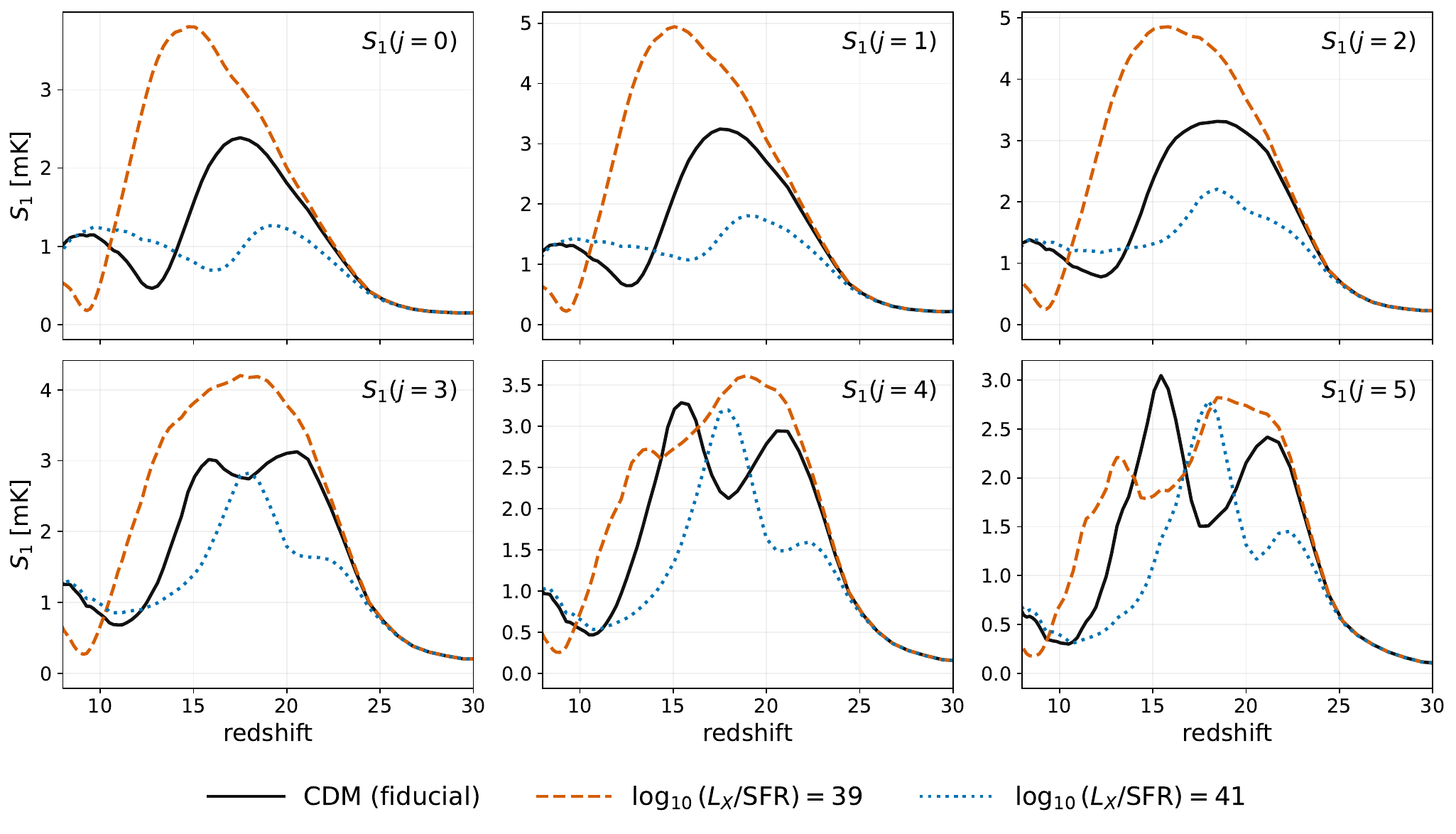}
    \caption{First-order $S_1(j)$ curves from the CDM $L_{\rm X}/{\rm SFR}$ sweep at logarithmic values 39 and 41 in ${\rm erg}\,{\rm s}^{-1}\,\mathrm{keV}^{-1}\,M_\odot^{-1}\,{\rm yr}$.}
    \label{fig:wst_1st_astro_lx}
\end{figure*}

\begin{figure*}
    \centering
    \includegraphics[width=0.9\hsize]{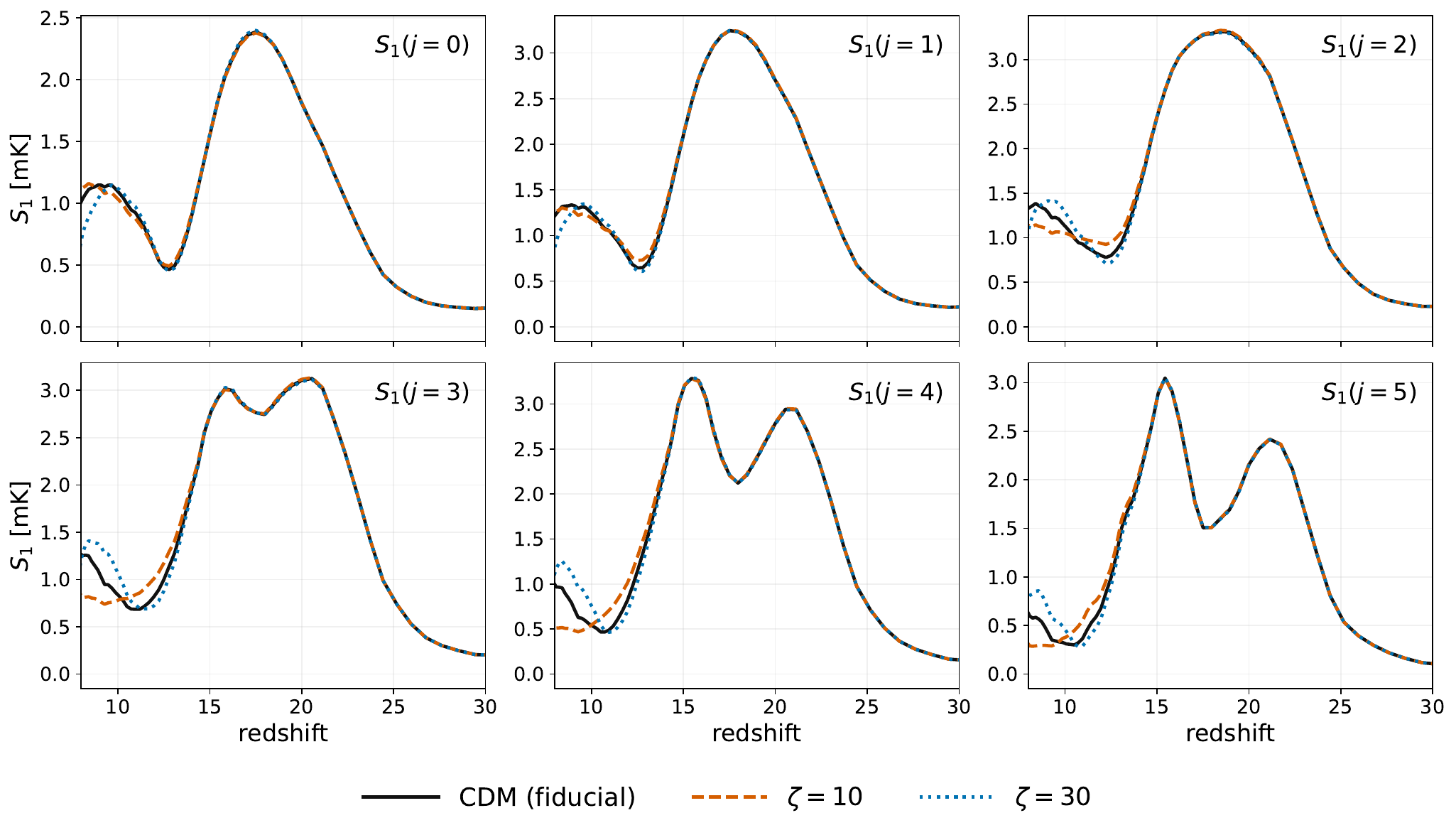}
    \caption{First-order $S_1(j)$ curves from the CDM ionizing-efficiency sweep at $\zeta=10$ and 30.}
    \label{fig:wst_1st_astro_zeta}
\end{figure*}

\subsubsection{Second-Order WST Coefficient}

Figure~\ref{fig:wst_2nd} shows the ordered ratio
$R(j_1,j_2)=S_2(j_1,j_2)/S_1(j_1)$. Several low-order pairs,
especially those anchored at $j_1=0$, show displaced peaks for
light FDM near the heating and early-reionization epochs. This
response arises indirectly because delayed source formation alters
the radiative backgrounds and consequently reorganizes the
cross-scale morphology of the 21~cm field. Because the denominator
can evolve more strongly than the numerator, $R$ can increase even
when the absolute fine-band amplitude decreases. At larger scales,
the differences between the models become weaker and are expressed
mainly as shifts in the timing of the characteristic features. By
normalizing $S_2$ with $S_1(j_1)$, $R$ suppresses the dependence on
the reference-band amplitude and provides a compact measure of
higher-order cross-scale morphology.

\begin{figure*}
    \centering
    \includegraphics[width=0.90\hsize]{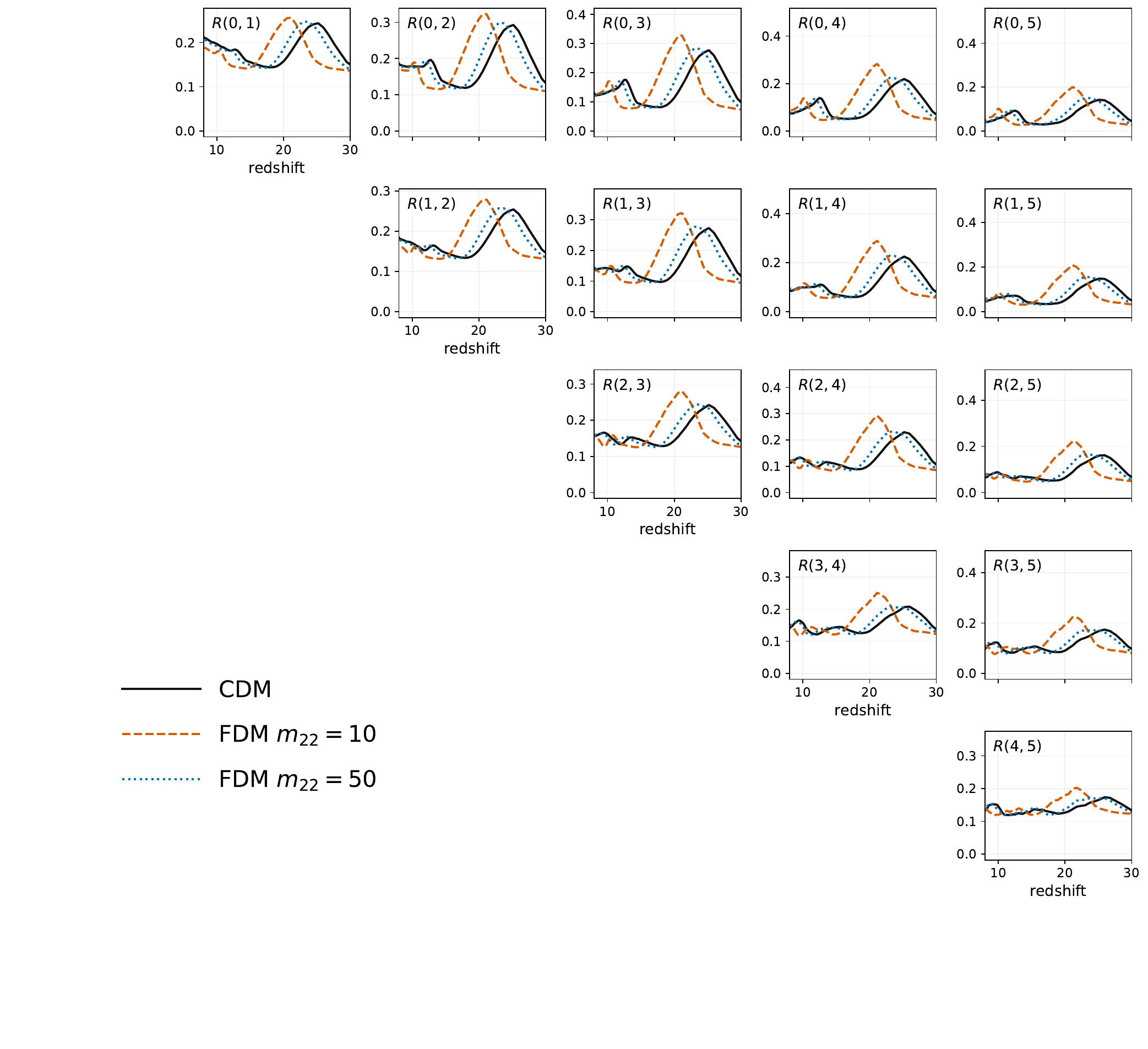}
    \caption{
    Second-order wavelet-scattering ratio $R=S_2/S_1$ without thermal noise.
    Dependence on the FDM mass parameter, comparing CDM with FDM models $m_{22}=10$ and $m_{22}=50$. The dependence on the HMF-shape parameter $\alpha$ is shown in Appendix~\ref{app:alpha_robustness}.
    Coefficients are selected from path metadata after excluding $S_0$.
    }
    \label{fig:wst_2nd}
\end{figure*}

Figures~\ref{fig:wst_2nd_astro_tvir} through \ref{fig:wst_2nd_astro_zeta} show that the same timing changes propagate into $R$: larger $T_{\rm vir}$ shifts the coupling features later, larger $L_\mathrm{X}/{\rm SFR}$ shifts and damps the heating-driven features, and larger $\zeta$ suppresses the low-redshift reionization response. The dependence on $\alpha$ is shown in Appendix~\ref{app:alpha_robustness}.

\begin{figure*}
    \centering
    \includegraphics[width=0.9\hsize]{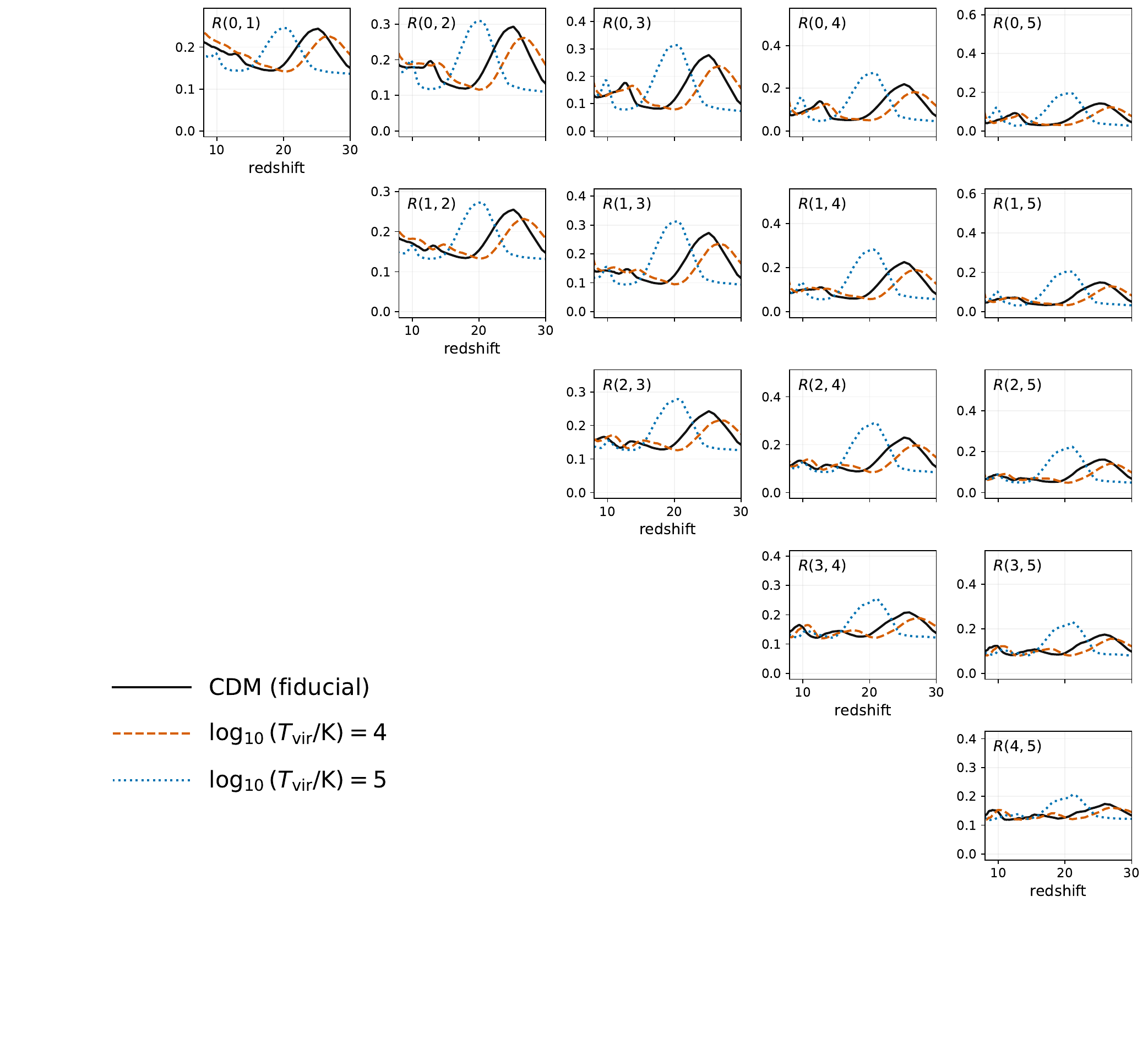}
    \caption{Second-order $R=S_2/S_1(j_1)$ curves from the CDM $T_{\rm vir}$ sweep, labeled $\log_{10}(T_{\mathrm{vir}}/{\rm K})=4$ and 5. The sweep is descriptive only.}
    \label{fig:wst_2nd_astro_tvir}
\end{figure*}

\begin{figure*}
    \centering
    \includegraphics[width=0.9\hsize]{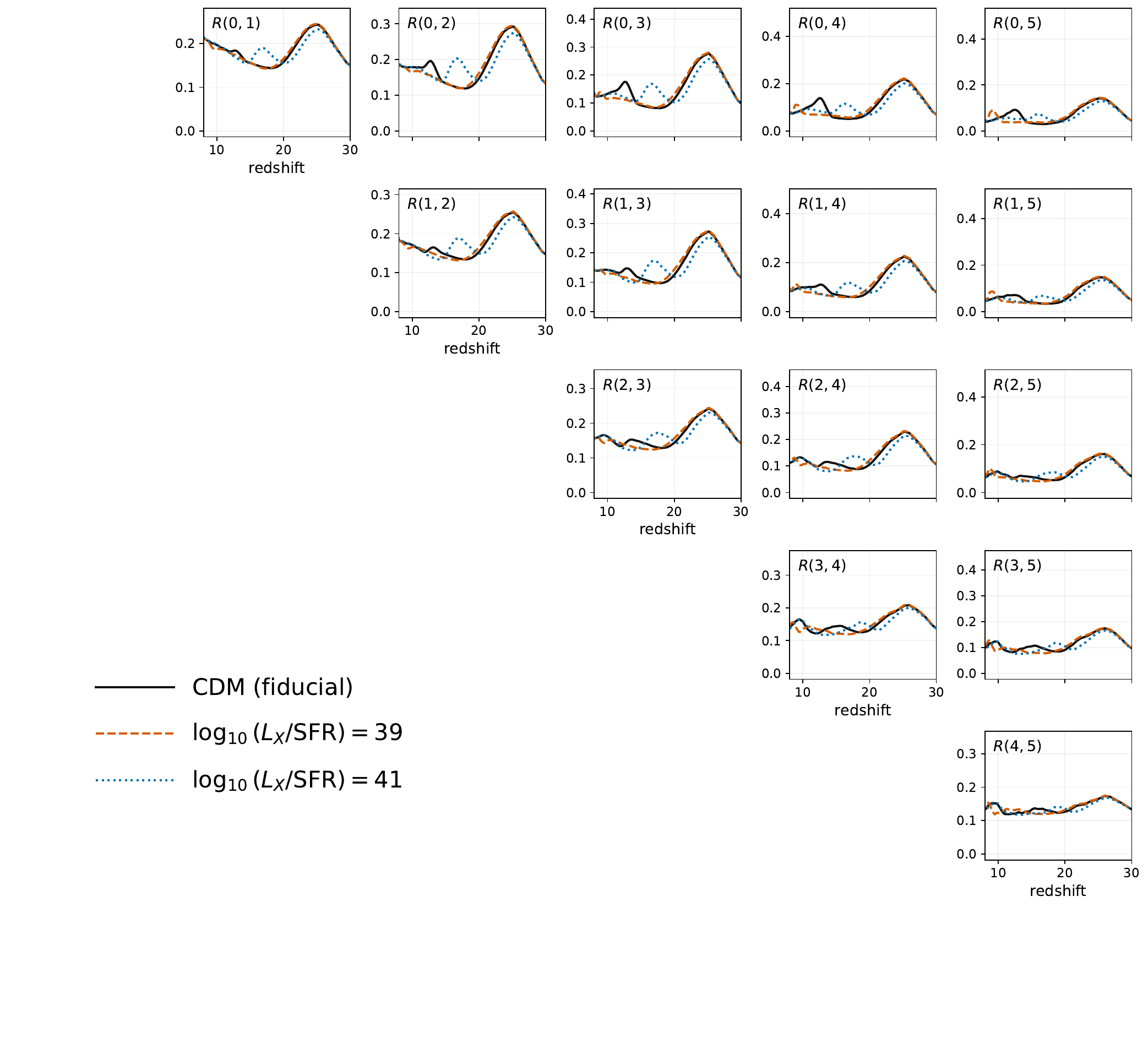}
    \caption{Second-order $R=S_2/S_1(j_1)$ curves from the CDM $L_{\rm X}/{\rm SFR}$ sweep at logarithmic values 39 and 41 in ${\rm erg}\,{\rm s}^{-1}\,\mathrm{keV}^{-1}\,M_\odot^{-1}\,{\rm yr}$.}
    \label{fig:wst_2nd_astro_lx}
\end{figure*}

\begin{figure*}
    \centering
    \includegraphics[width=0.9\hsize]{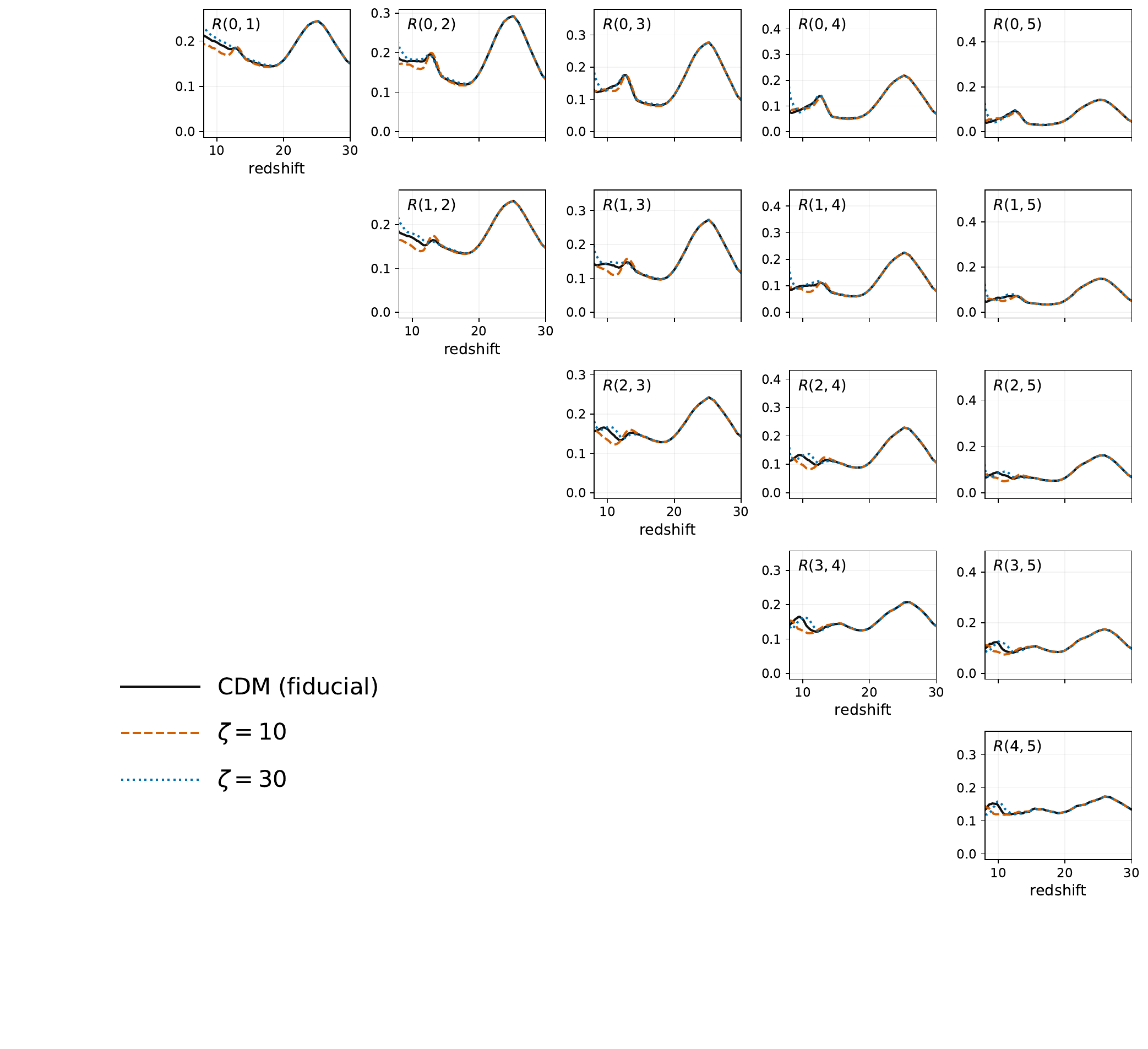}
    \caption{Second-order $R=S_2/S_1(j_1)$ curves from the CDM ionizing-efficiency sweep at $\zeta=10$ and 30.}
    \label{fig:wst_2nd_astro_zeta}
\end{figure*}

\subsection{Observational Requirements and the Controlled Instrumental Transfer Function}
\label{subsec:instrument}

We model the instrumental response using a controlled analytic map
operator inspired by SKA1-Low. The operator is intentionally defined
independently of the released AA*, AA4, and core-selection station
layouts \citep{SKA1LowConfig}, allowing the effects of finite Fourier
coverage and thermal noise to be isolated within a common map-space
framework for the PS and WST. It assumes 256 stations with a diameter
of 38\,m, a nominal baseline range of 40\,m to 5\,km, an integration
time of 1000\,hr, two polarizations, and a bandwidth of 2\,MHz. The
present implementation represents the static Fourier-sampling and
radiometer-noise stage of the observation model. A full
visibility-domain treatment additionally incorporates Earth-rotation
synthesis, direction-dependent beam responses, deconvolution, and
calibration systematics.

Each transverse map has side length
$L_\perp=300\,{\rm Mpc}$ and $N=200$ pixels.
To avoid confusion with the WST ratio $R=S_2/S_1$, we denote the fractional Nyquist radius by $\varrho$.
For the Fourier index $\boldsymbol q=(q_x,q_y)$, we define
\begin{equation}
k_\perp(\boldsymbol q)
=
\frac{2\pi}{L_\perp}
\sqrt{q_x^2+q_y^2},
\qquad
\varrho(\boldsymbol q)
=
\frac{k_\perp(\boldsymbol q)}{k_{\rm Nyq}}
=
\frac{2}{N}\sqrt{q_x^2+q_y^2},
\label{eq:normalized_fourier_radius}
\end{equation}
where $\boldsymbol q=(q_x,q_y)$ is the discrete transverse Fourier index, $L_\perp$ is the transverse comoving map size, $N$ is the number of pixels per side, $k_\perp$ is the transverse comoving wavenumber, and $k_{\rm Nyq}=\pi N/L_\perp$ is the Nyquist wavenumber. The circular Fourier boundary is set at $\varrho=1$.

To define the Fourier support of the controlled map operator, we
adopt the station diameter as the characteristic angular scale.
We use $D_{\rm st}=38\,{\rm m}$, consistent with the station scale
adopted in earlier SKA1-Low design studies, and approximate the
station-beam FWHM by that of an idealized circular aperture,
$\theta_{\rm PB}=1.02\lambda/D_{\rm st}$. The angular width of the
$N=200$ transverse map is set equal to this FWHM. This choice fixes
the Fourier-cell spacing to $\Delta u=1/\theta_{\rm PB}$ and provides
a common normalized map operator for the matched PS and WST
comparison.

A baseline $b$ then maps to the radial grid coordinate
\begin{equation}
\varrho_b
=
\frac{2(1.02)b}{ND_{\rm st}}.
\end{equation}
Here $b$ is the baseline length and $D_{\rm st}$ is the station diameter. We adopt $b_{\min}=40\,{\rm m}$ as a station-diameter-scale lower
cutoff, which gives $\varrho_{\min}=0.0107$. The upper support is set by
the Nyquist boundary of the simulated map, $\varrho_{\max}=1$. Under the
adopted field-of-view mapping, this boundary corresponds to
$b_{\rm Nyq}\simeq3.73\,{\rm km}$. Baselines longer than this scale
are not represented on the map grid and therefore do not alter the
instrumental transfer function.

The wavelength cancels in this normalized-grid prescription, so
the same Fourier mask is applied at every redshift. A physical
fixed array would instead sample
$k_\perp=2\pi b/[\lambda D_M(z)]$ and would have
redshift-dependent comoving coverage. The present operator is
therefore a controlled map-space approximation to sparse
interferometric sampling.

The instrumental amplitude transfer function is
\begin{equation}
\begin{aligned}
T_{\rm inst}(\boldsymbol q)
&=
M_{\rm ann}(\boldsymbol q)
M_{\rm occ}(\boldsymbol q)
W[\varrho(\boldsymbol q)],\\
M_{\rm ann}(\boldsymbol q)
&=
\Theta\!\left[\varrho(\boldsymbol q)-\varrho_{\min}\right]
\Theta\!\left[\varrho_{\max}-\varrho(\boldsymbol q)\right],\\
W(\varrho)
&=
\exp\left[-\left(\frac{\varrho}{0.5}\right)^4\right].
\end{aligned}
\label{eq:instrument_transfer}
\end{equation}
Here $M_{\rm ann}$ is the annular Fourier-support mask, $M_{\rm occ}$ is the incomplete-occupancy mask, $W$ is the radial taper, $\Theta$ denotes the Heaviside step function, and $T_{\rm inst}$ denotes the instrumental map transfer function. This notation distinguishes it from the FDM linear-matter transfer function in Eq.~(\ref{eq:fdm_transfer_redux}).
The binary mask $M_{\rm occ}$ retains 30\% of the Fourier modes
within the annulus and represents incomplete $uv$ coverage. The
super-Gaussian taper leaves the inner Fourier region largely
unchanged and produces a smooth high-$k$ roll-off near the outer
boundary. This reduces ringing associated with a sharp Fourier
cutoff. The occupancy fraction and taper parameters are
phenomenological components of the controlled operator and are
not calibrated to a released SKA1-Low baseline-density model.

For a mean-subtracted 2\,MHz map, the transferred and noisy
products are
\begin{equation}
\begin{aligned}
\widetilde I_{\rm tr}(\boldsymbol q,z)
&=
T_{\rm inst}(\boldsymbol q)
\widetilde I_{\rm clean}(\boldsymbol q,z),\\
\widetilde I_{\rm obs}(\boldsymbol q,z)
&=
T_{\rm inst}(\boldsymbol q)
\left[
\widetilde I_{\rm clean}(\boldsymbol q,z)
+
\widetilde n_0(\boldsymbol q,z)
\right].
\end{aligned}
\label{eq:mock_observed_map}
\end{equation}
Here a tilde denotes a two-dimensional Fourier transform, $I_{\rm clean}$ is the clean mean-subtracted map, $I_{\rm tr}$ is the map after applying $T_{\rm inst}$, $I_{\rm obs}$ is the transferred map including thermal noise, and $n_0$ is the pre-transfer thermal-noise field. We use $S_1^{\rm clean}$ and $S_1^{\rm tr}$ for first-order WST coefficients measured from $I_{\rm clean}$ and $I_{\rm tr}$, respectively. The signal and thermal noise therefore occupy the same retained
Fourier modes.

The thermal-noise level is set using
\begin{equation}
T_{\rm sys}
=
1.1T_{\rm sky}+40\,{\rm K},
\qquad
T_{\rm sky}
=
60
\left(\frac{\nu}{300\,{\rm MHz}}\right)^{-2.55}
{\rm K},
\label{eq:system_temperature}
\end{equation}
Here $T_{\rm sys}$ is the system temperature, $T_{\rm sky}$ is the sky-temperature contribution, and $\nu$ is the observing frequency. We additionally use the radiometer scaling for the adopted number of stations, integration time, bandwidth, and polarizations.
For each
redshift slab, the Fourier-space noise field is normalized so that,
after multiplication by the instrumental transfer function and inverse Fourier
transformation, its image-domain rms matches the value implied by
these observing parameters. This procedure places the thermal noise
on the same retained Fourier modes as the signal.

We next isolate the scale-dependent attenuation of the signal caused
by the instrumental transfer function. For a first-order wavelet response,
\begin{equation}
I_{\rm tr}*\psi_{j,\ell}
=
\mathcal{F}^{-1}
\left[
T_{\rm inst}(\boldsymbol k)
\widetilde I_{\rm clean}(\boldsymbol k)
\widetilde\psi_{j,\ell}(\boldsymbol k)
\right].
\end{equation}
Here $\mathcal{F}^{-1}$ denotes the inverse Fourier transform of the quantity in square brackets. The symbol $\boldsymbol k$ is the transverse Fourier wavevector, and $\widetilde\psi_{j,\ell}$ is the Fourier transform of the Morlet wavelet.
Because each Morlet wavelet has a finite Fourier-space passband,
the attenuation of $S_1(j)$ is determined by the instrumental transfer over the
full range of modes sampled by that band. The fine-scale wavelets
overlap the high-$k$ taper and the outer Fourier boundary, while the
broad-scale wavelets receive contributions from the sparsely sampled
low but nonzero modes. The incomplete Fourier occupancy contributes
to the attenuation across all wavelet scales.

To characterize this overlap, we define the orientation-averaged
wavelet power
\begin{equation}
\Phi_j(\boldsymbol k)
=
\frac{1}{L}
\sum_{\ell=0}^{L-1}
\left|
\widetilde\psi_{j,\ell}(\boldsymbol k)
\right|^2.
\end{equation}
Here $\Phi_j(\boldsymbol k)$ is the orientation-averaged Fourier-space wavelet power at scale $j$.
We then define the corresponding retained-amplitude proxy
\begin{equation}
A_j^{\rm filt}
=
\left[
\frac{
\sum_{\boldsymbol k}
\Phi_j(\boldsymbol k)
\overline{|T_{\rm inst}|^2}(k)
}{
\sum_{\boldsymbol k}
\Phi_j(\boldsymbol k)
}
\right]^{1/2},
\label{eq:s1_attenuation_proxy}
\end{equation}
where $\overline{|T_{\rm inst}|^2}(k)$ is the azimuthally averaged transfer
power. The quantity $A_j^{\rm filt}$ measures the fraction of
wavelet-band amplitude transmitted by the map operator under a
flat-spectrum Gaussian approximation. It therefore provides a
transfer-only reference for the scale dependence of
$S_1^{\rm tr}/S_1^{\rm clean}$.

Figure~\ref{fig:transfer} provides a scale-by-scale view of how the
analytic instrumental transfer operator modifies the first-order WST coefficients.
Panel (a) compares the orientation-averaged Morlet passbands
$\Phi_j(k)$ with the azimuthally averaged instrumental transfer power
$\overline{|T_{\rm inst}|^2}(k)$. The fine-scale bands at small $j$ extend into
the high-$k$ region suppressed by the radial taper, while the broadest
bands approach the low-$k$ region affected by the inner cutoff and
sparse Fourier sampling. The intermediate bands overlap most strongly
with the well-transmitted part of Fourier space.

Panel (b) compares the retained-amplitude proxy $A_j^{\rm filt}$ with
the attenuation measured directly from the transferred maps,
$S_1^{\rm tr}(j)/S_1^{\rm clean}(j)$. The points show the median over
the CDM, $m_{22}=10$, and $m_{22}=50$ models at $z\simeq10$, 15,
and 20, while the intervals show the corresponding 16th to 84th
percentile range. The measured retained amplitude increases from the
finest band to a maximum near $j=2$ and then decreases toward the
broadest band. This scale dependence follows the overlap between each
Morlet passband and the transmitted Fourier region. The proxy captures
this overall ordering and agrees closely with the measured attenuation
at intermediate and broad scales. At the finest scales, the measured
retained amplitude is larger than the flat-spectrum estimate because
the actual coefficient also weights the nonuniform 21\,cm power within
the passband and includes the nonlinear modulus operation defining
$S_1$.

The PS and WST are computed from the same transferred maps, retained
Fourier modes, integration time, bandwidth, and paired thermal-noise
realizations. Their Fisher information is therefore compared under a
single observational operator, allowing differences between the
constraints to be attributed to the distinct scale and morphological
information summarized by the two statistics. A real-data
implementation will replace the analytic map operator with a
visibility-domain response incorporating the station beam, calibration,
visibility gridding, deconvolution, mapmaking, and empirical
instrumental-transfer-function characterization.

\begin{figure*}[htbp]
    \centering
    \begin{minipage}[t]{0.49\textwidth}
      \centering
      \includegraphics[width=\linewidth]
      {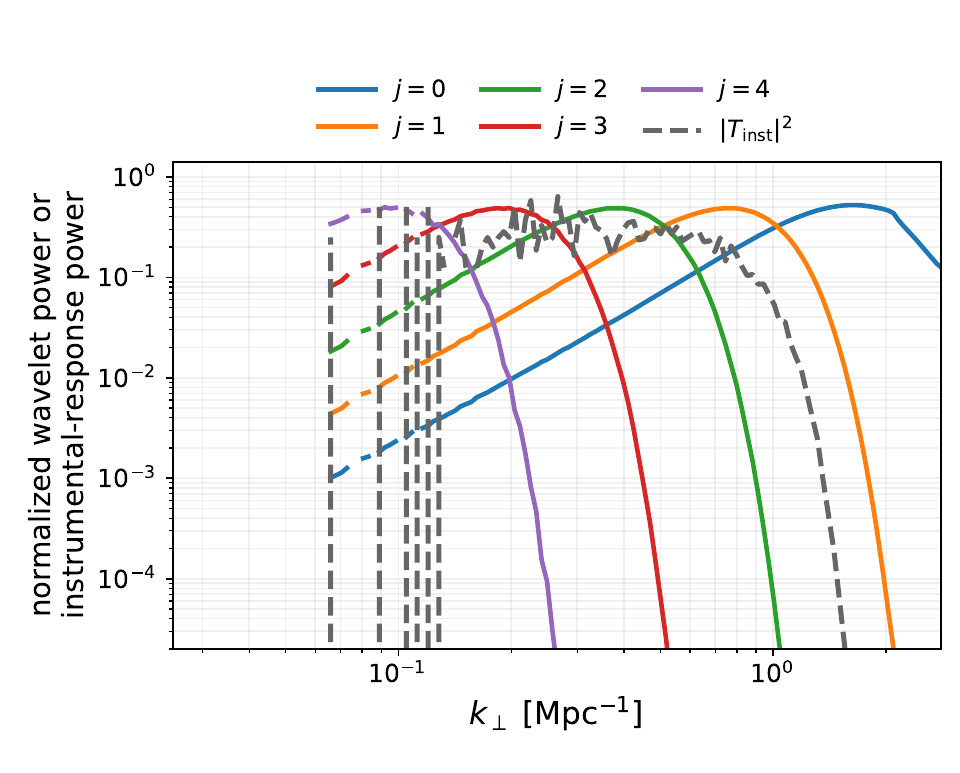}
      \\[-0.4ex]
      {\small\bfseries (a)}
    \end{minipage}
    \hfill
    \begin{minipage}[t]{0.49\textwidth}
      \centering
      \includegraphics[width=\linewidth]
      {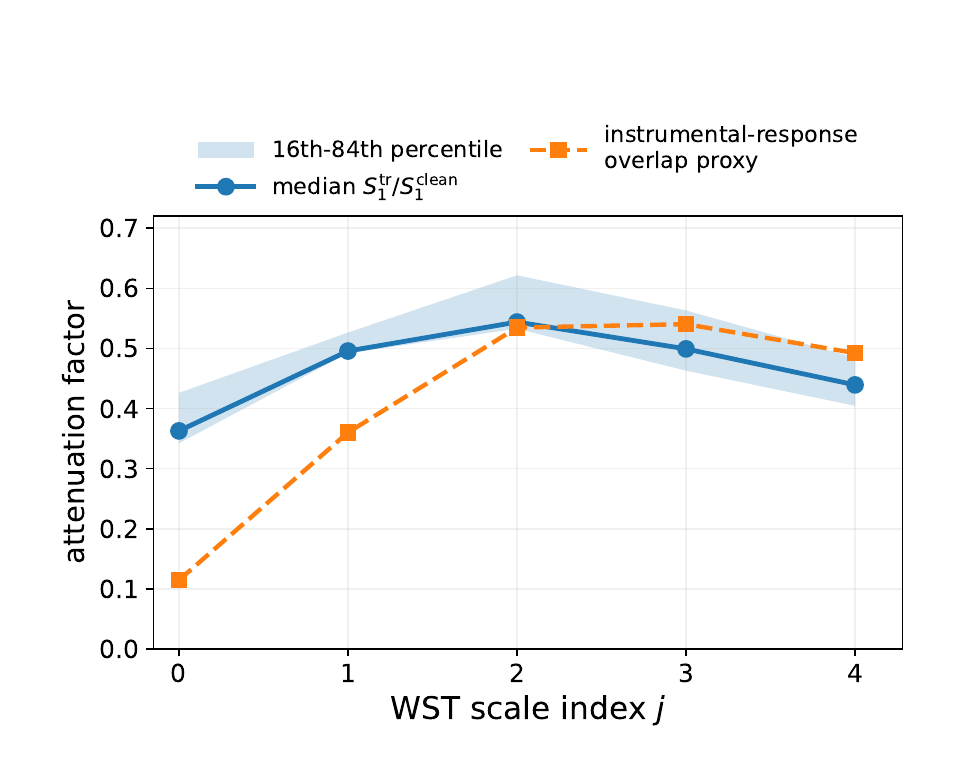}
      \\[-0.4ex]
      {\small\bfseries (b)}
    \end{minipage}

   \caption{Scale-dependent response of the first-order WST coefficients to the controlled analytic instrumental map operator. Panel (a) shows the orientation-averaged Morlet power passbands for $j=0,\ldots,4$ together with the azimuthally averaged instrumental transfer power $\overline{|T_{\rm inst}|^2}$. Smaller $j$ probes finer, higher-$k$ structures, while larger $j$ shifts the response toward broader, lower-$k$ structures. Panel (b) compares the directly measured transfer-only attenuation $S_1^{\rm tr}(j)/S_1^{\rm clean}(j)$ with the flat-spectrum wavelet-transfer overlap proxy $A_j^{\rm filt}$. Here $S_1^{\rm tr}$ and $S_1^{\rm clean}$ are measured from the transferred and clean maps, respectively. The points show the median and the shaded region the 16th to 84th percentile range over three dark-matter models and $z\simeq10$, 15, and 20. Intermediate scales retain the largest fraction of their amplitude because their passbands overlap most strongly with the transmitted Fourier region.}
    \label{fig:transfer}
\end{figure*}

\subsection{\texorpdfstring{
Separation of Instrumental Transfer and Thermal Noise
}{
Separation of Instrumental Transfer and Thermal Noise
}}
\label{subsec:mock_stages}

Figures~\ref{fig:wst_1st} through
\ref{fig:wst_2nd_astro_zeta} characterize the intrinsic
scale-dependent and redshift-dependent responses of the full
first-order and second-order coefficient sets.
Figure~\ref{fig:wst_stages} follows these model differences
through the controlled instrumental transfer and thermal-noise
model.

Panel (a) shows a representative mean-subtracted
$m_{22}=10$ map constructed from a 2\,MHz frequency interval
at $z\simeq15$. The three maps are displayed over the same
temperature range. The clean map contains the simulated
21\,cm field and has an rms of 6.2\,mK. Applying the Fourier
support, sparse mode occupancy, and radial taper reduces the rms
to 3.1\,mK and weakens the coherent image contrast. Adding one
thermal-noise realization increases the rms to 31.0\,mK and
produces the fine-grained fluctuations visible in the final map.
This sequence separates the systematic attenuation of the
21\,cm structure by the instrumental transfer function from the additional
small-scale fluctuations introduced by thermal noise.

Panel (b) tracks how much of the intrinsic FDM-to-CDM contrast
survives in three representative WST components:
$S_1(j=2)$, $R(0,1)$, and $R(0,2)$. The coefficient
$S_1(j=2)$ samples an intermediate wavelet band that overlaps
strongly with the transmitted Fourier region, while $R(0,1)$
and $R(0,2)$ probe low-order cross-scale modulation with
prominent intrinsic FDM responses. For each component, the
vertical axis gives the FDM coefficient divided by the CDM
coefficient at the same redshift and observational stage.
A ratio above unity means that the FDM coefficient is larger
than the corresponding CDM value, while a ratio below unity
means that it is smaller. The distance from unity quantifies
the remaining model contrast in that individual component.

The ratios are shown at $z\simeq10$, 15, and 20 for the clean,
transfer-only, and transfer-plus-noise stages. The clean ratios
measure the intrinsic differences between the simulated FDM and
CDM fields. The transfer-only ratios show how finite Fourier
coverage, sparse mode occupancy, and the radial taper modify
these differences. The transfer-plus-noise ratios show the
contrast remaining after the addition of thermal noise.

For the transfer-plus-noise stage, the markers show the median
FDM-to-CDM ratio from 20 paired thermal-noise realizations, and
the vertical intervals show the corresponding 16th to 84th
percentile range. Each FDM map and its CDM counterpart receive
the same noise realization before the ratio is formed. This
pairing removes much of the scatter associated with unrelated
noise draws and isolates the response of the model contrast to
the thermal-noise contribution.

The clean coefficients show substantial departures from unity
at several redshifts, particularly for the low-order ratios near
$z\simeq20$. The instrumental transfer function reduces some of these
departures, and the addition of thermal noise generally moves
the ratios further toward unity while increasing their scatter.
Thus, under the adopted mock-observation conditions, these
representative components become increasingly noise dominated
and their individual FDM-to-CDM separation is weakened. The
degree of degradation depends on the coefficient, FDM mass,
and redshift.

Figure~\ref{fig:wst_stages} therefore provides a component-level
view of how selected WST responses propagate through the
observational treatment. Each individual coefficient captures
only a particular combination of spatial scale and redshift and
therefore provides partial parameter sensitivity when considered
on its own. The Fisher analysis in Section~\ref{sec:fisher}
combines five $S_1$ coefficients and ten $R$ coefficients in
each of the 17 selected redshift slabs, yielding 85 first-order
and 170 second-order WST components. Their joint scale-dependent
and redshift-dependent evolution forms a characteristic response
pattern for each model parameter. Combining these complementary
responses with their full covariance substantially strengthens
the parameter constraints relative to those obtained from any
single WST component. The quantitative comparison therefore
uses the complete covariance-weighted multiscale and
multiredshift information after applying the common transfer and
thermal-noise treatment.

\begin{figure*}[htbp]
    \centering

    \includegraphics[width=0.94\textwidth]
    {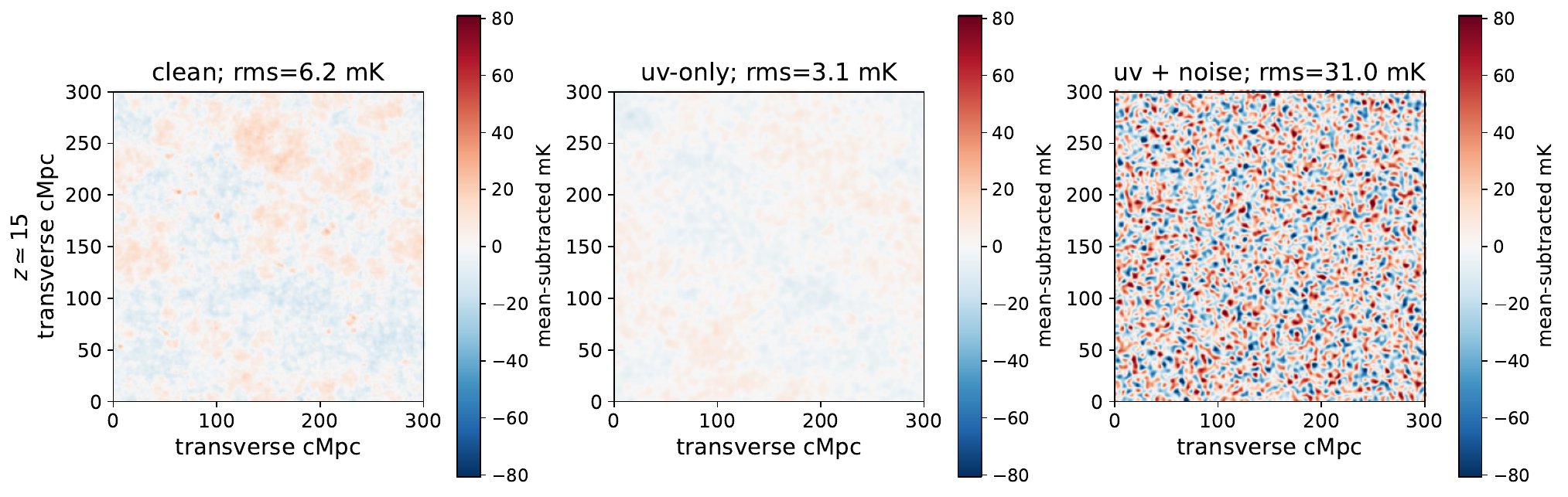}
    \\[-0.2ex]
    {\small\bfseries (a)}

    \vspace{0.8ex}

    \includegraphics[width=0.96\textwidth]
    {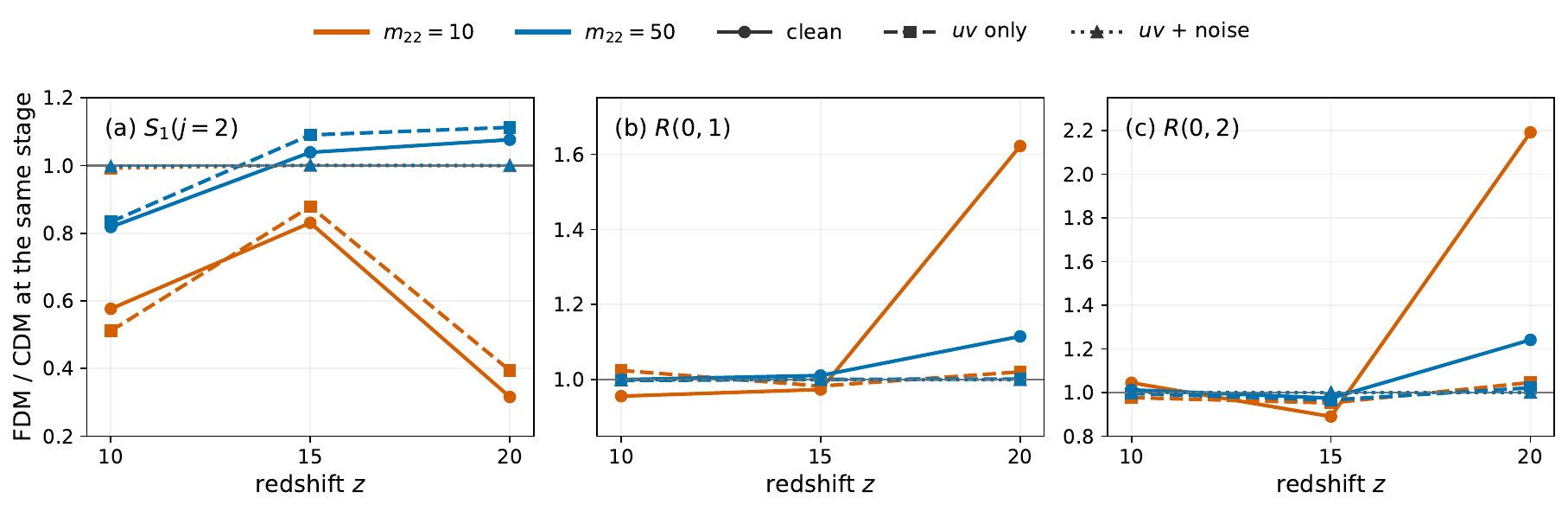}
    \\[-0.2ex]
    {\small\bfseries (b)}

    \caption{
    Clean, transfer-only, and transfer-plus-noise stages.
    Panel (a) shows a representative mean-subtracted
    $m_{22}=10$ map at $z\simeq15$ for a 2\,MHz frequency
    interval. Panel (b) shows the FDM-to-CDM ratios of
    $S_1(j=2)$, $R(0,1)$, and $R(0,2)$ at
    $z\simeq10$, 15, and 20. For the noisy stage, markers
    and intervals show the median and 16th to 84th percentiles
    from 20 paired thermal-noise realizations.
    }
    \label{fig:wst_stages}
\end{figure*}

\subsection{Idealized Foreground-Wedge Response}
\label{subsec:wedge}

We test the response to idealized foreground avoidance using
8\,MHz light-cone chunks centered at nine redshifts between
$z=10$ and 20. Each chunk is Fourier transformed along the
transverse and line-of-sight directions, modes within the
foreground wedge are set to zero, and the filtered field is
inverse transformed. We then average the central 2\,MHz portion
along the line of sight to construct the two-dimensional map
used for the WST measurement.

The foreground-wedge boundary is defined by
\begin{equation}
|k_\parallel|
<
s_{\rm w}(z)\sin\theta\,k_\perp+b,
\qquad
s_{\rm w}(z)
=
\frac{D_M(z)H(z)}{c(1+z)}.
\label{eq:wedge_boundary}
\end{equation}
Here $s_{\rm w}(z)$ is the foreground-wedge slope in comoving Fourier space, $k_\parallel$ and $k_\perp$ are the line-of-sight and transverse comoving wavenumbers, $D_M(z)$ is the transverse comoving distance, $H(z)$ is the Hubble parameter, $c$ is the speed of light, $\theta$ sets the angular extent of the wedge, and $b$ is an additive buffer in $k_\parallel$. Both positive and negative $k_\parallel$ modes satisfying this
condition are removed. We consider three idealized cases: an
optimistic wedge with
$(\sin\theta,b)=(0.5,0)$,
a horizon wedge with
$(\sin\theta,b)=(1,0)$,
and a buffered horizon wedge with
$(\sin\theta,b)=(1,0.1\,{\rm Mpc}^{-1})$.

These masks remove the selected Fourier modes perfectly and do
not add residual foreground power. The calculation is therefore
a coefficient-level mode-removal diagnostic rather than a
realistic foreground-subtraction or foreground-avoidance
forecast. The wedge-filtered summaries are not included in the
Fisher analysis.

For each block
$B\in\{S_1,R,{\rm WST}\}$,
we quantify the absolute fractional coefficient response as
\begin{equation}
\chi_B
=
{\rm median}_{a,\mathcal{D},z}
\left[
\frac{
\left|
d_{B,a}^{\rm wedge}(\mathcal{D},z)
-
d_{B,a}^{\rm none}(\mathcal{D},z)
\right|
}{
\left|
d_{B,a}^{\rm none}(\mathcal{D},z)
\right|
}
\right].
\label{eq:wedge_response_metric}
\end{equation}
Here $a$ indexes the coefficients within block $B$, $\mathcal{D}$ indexes the simulated dark-matter models, and $z$ indexes the selected redshift centers. The symbol $\mathcal{D}$ is used for the model index to avoid confusion with the region mass $M$ in Eq.~(\ref{eq:fdm_env}). For the combined WST block,
$\boldsymbol d_{\rm WST}$ is formed by concatenating the five
$S_1$ coefficients and ten $R$ coefficients.

Figure~\ref{fig:foreground_wst_response} presents the responses
for all three wedge definitions. The median fractional change
in $S_1$ is close to order unity, whereas the median response
of $R$ lies between 0.17 and 0.26 across the three masks.
Thus, at the coefficient level, the normalized second-order
ratios are less strongly reshaped by idealized wedge filtering
than the first-order amplitudes.

The smaller coefficient-level response of $R$ indicates that the
normalized second-order structure is more stable under idealized
foreground avoidance than the first-order amplitudes. The vertical
intervals show the 16th to 84th percentile range over coefficients,
models, and redshifts, revealing how this stability varies across
scale pairs and evolutionary stages. These results identify the
$R$ block as a promising component of a foreground-aware WST
analysis. A natural next step is to combine the wedge-filtered WST
vector with a covariance model including foreground residuals and
quantify the FDM information retained across scales and redshifts.

\begin{figure*}[htbp]
    \centering
    \includegraphics[width=0.88\textwidth]
    {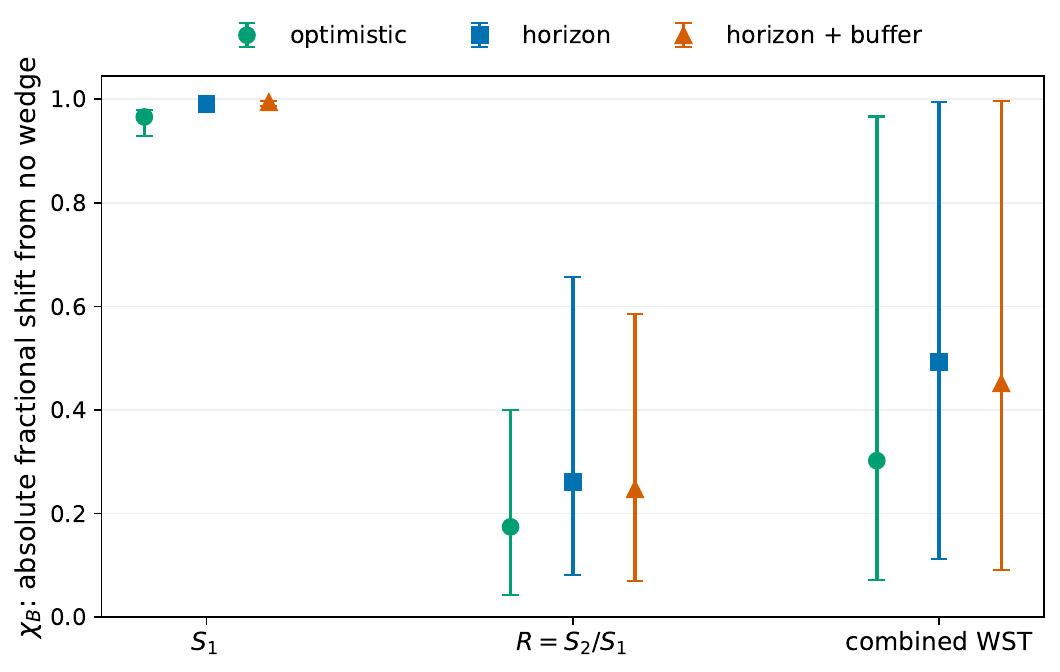}

    \caption{
    Coefficient-level response to the three idealized
    foreground-wedge definitions for the $S_1$ block,
    the normalized second-order block $R=S_2/S_1$, and their
    combined WST vector. For each wedge definition, the points
    show the median absolute fractional response, while the
    vertical intervals show the 16th to 84th percentile range
    over the selected coefficients, dark-matter models, and
    redshift centers. The intervals quantify heterogeneity in
    the coefficient response and are not parameter
    uncertainties. A smaller response does not by itself imply
    that FDM-to-CDM separation or Fisher information is retained
    after foreground avoidance.
    }
    \label{fig:foreground_wst_response}
\end{figure*}

\section{Fisher Comparison with the Power Spectrum}
\label{sec:fisher}

We use the Fisher-matrix formalism to compare the local
information carried by the matched two-dimensional PS and WST
summaries. All statistic choices are constructed from the same
redshift slabs and use the same instrumental transfer,
thermal-noise prescription, and parameter-response simulations.
A separate regularized covariance is estimated for each data-vector
choice. The purpose of this calculation is to determine whether the
WST contributes information complementary to the selected
two-dimensional PS under identical mock-observation assumptions.

For a model parameter vector $\boldsymbol{\theta}$, we denote the
expectation value of the data vector by
$\boldsymbol{\mu}(\boldsymbol{\theta})=\langle\mathbf{d}\rangle$.
Assuming a local multivariate-Gaussian likelihood and neglecting the
parameter dependence of the covariance, the Fisher matrix is defined
as
\begin{equation}
F_{ij}
=
\left(\frac{\partial\boldsymbol{\mu}}{\partial\theta_i}\right)^{T}
\mathbf{C}^{-1}
\left(\frac{\partial\boldsymbol{\mu}}{\partial\theta_j}\right).
\end{equation}
Here $F_{ij}$ is the Fisher matrix, $\boldsymbol{\mu}$ is the mean data vector, $\theta_i$ and $\theta_j$ are model parameters, the superscript $T$ denotes vector transpose, and $\mathbf C$ is the data covariance matrix. The derivatives and covariance are evaluated at the fiducial
parameter point $\boldsymbol{\theta}_{\rm fid}$. Within this local
Gaussian approximation, the inverse Fisher matrix gives the
marginalized parameter covariance. The marginalized uncertainty on
parameter $\theta_i$ is
\begin{equation}
\sigma(\theta_i)
=
\sqrt{\left(\mathbf{F}^{-1}\right)_{ii}}.
\end{equation}
Here $\mathbf F^{-1}$ is the inverse Fisher matrix and $\sigma(\theta_i)$ is the marginalized standard deviation of $\theta_i$.

\subsection{Fisher Setup}
\label{subsec:fisher_setup}

The primary Fisher comparison is performed in the
three-parameter space
\begin{equation}
{\color{black}
\boldsymbol{\theta}
=
\left(m_{22},\,\ell_\mathrm{X},\,\zeta\right),
}
\end{equation}
where
\begin{equation}
{\color{black}
\ell_\mathrm{X}
\equiv
\log_{10}
\left[
\frac{L_\mathrm{X}/{\rm SFR}}
{{\rm erg}\,{\rm s}^{-1}\,\mathrm{keV}^{-1}\,
(M_\odot\,{\rm yr}^{-1})^{-1}}
\right].
}
\end{equation}
The fiducial parameter point is
$\boldsymbol{\theta}_{\rm fid}=(10,40.00,20)$.
The cosmological parameters are held fixed throughout the
comparison.

The primary Fisher analysis is defined as a controlled comparison
of summary statistics within a fixed atomic-cooling source
prescription. We hold
$T_{\rm vir}=2\times10^4\,{\rm K}$ fixed as the minimum virial
temperature of source-hosting halos and vary
$(m_{22},\ell_\mathrm{X},\zeta)$. These three parameters describe the
suppression of early structure formation, the effective X-ray
heating normalization, and the ionizing efficiency while preserving
the same source-host population across the PS and WST calculations.
This setup isolates the scale-dependent and redshift-dependent
information contributed by the two summary statistics within a
common source model.

For each redshift slab $z_k$, the common parent data vector is
\begin{equation}
{\color{black}
\mathbf{d}(z_k)
=
\left(P(z_k),\,S_1(z_k),\,R(z_k)\right).
}
\end{equation}

Here $z_k$ denotes the center of the $k$th redshift slab and $\mathbf d(z_k)$ is the summary-statistic data vector for that slab. $P$ contains five bins of the two-dimensional PS
measured from the same map used for the WST calculation. The
first-order WST block contains five orientation-averaged
coefficients $S_1$, while the second-order block contains ten
ratios
$R(j_1,j_2)=S_2(j_1,j_2)/S_1(j_1)$
with $j_1<j_2$. The zeroth-order coefficient $S_0$ is excluded
from the Fisher data vector. The primary analysis uses 17
nonoverlapping 2\,MHz slabs over $8\leq z\leq25$. Three
additional slabs at $z>25$ are used only for the redshift-range
sensitivity test discussed in
Section~\ref{subsec:redshift_information}.

\begin{table*}[htbp]
\centering
\caption{Setup of the matched two-dimensional Fisher comparison}
\label{tab:fisher_setup}

\begingroup
\footnotesize
\renewcommand{\arraystretch}{1.35}
\setlength{\tabcolsep}{8pt}

\begin{tabular}{@{}ll@{}}
\hline

\parbox[t]{0.25\textwidth}{%
  \raggedright
  \textbf{Quantity}
}
&
\parbox[t]{0.67\textwidth}{%
  \raggedright
  \textbf{Adopted setup}
}
\\[0.8em]

\hline
\\[-0.4em]

\parbox[t]{0.25\textwidth}{%
  \raggedright
  Fiducial and varied parameters
}
&
\parbox[t]{0.67\textwidth}{%
  \raggedright
  $(m_{22},\ell_\mathrm{X},\zeta)=(10,40.00,20)$.
  The three parameters are varied locally, while the
  cosmological parameters are fixed.
}
\\[1.0em]

\parbox[t]{0.25\textwidth}{%
  \raggedright
  Redshift and map definition
}
&
\parbox[t]{0.67\textwidth}{%
  \raggedright
  Seventeen nonoverlapping 2\,MHz slabs over
  $8\leq z\leq25$. Each slab is represented by a
  $200^2$ transverse map.
}
\\[1.0em]

\parbox[t]{0.25\textwidth}{%
  \raggedright
  Summary statistics
}
&
\parbox[t]{0.67\textwidth}{%
  \raggedright
  Five two-dimensional PS bins, five $S_1$ coefficients,
  and ten ratios $R=S_2/S_1$ per slab.
}
\\[1.0em]

\parbox[t]{0.25\textwidth}{%
  \raggedright
  WST configuration
}
&
\parbox[t]{0.67\textwidth}{%
  \raggedright
  $J=5$, $L=4$, and \texttt{max\_order}$=2$.
  The zeroth-order coefficient $S_0$ is excluded from the
  Fisher data vector.
}
\\[1.0em]

\parbox[t]{0.25\textwidth}{%
  \raggedright
  Instrument model
}
&
\parbox[t]{0.67\textwidth}{%
  \raggedright
  A controlled analytic instrumental-transfer model with 256 stations
  of diameter 38\,m, baselines from 40\,m to 5\,km,
  an integration time of 1000\,hr, and a bandwidth of
  2\,MHz.
}
\\[1.0em]

\parbox[t]{0.25\textwidth}{%
  \raggedright
  Covariance
}
&
\parbox[t]{0.67\textwidth}{%
  \raggedright
  A total of 150 direct noisy-map realizations are used,
  with statistic-specific standardization and
  Ledoit-Wolf regularization.
}
\\[1.0em]

\parbox[t]{0.25\textwidth}{%
  \raggedright
  Principal limitation
}
&
\parbox[t]{0.67\textwidth}{%
  \raggedright
  The covariance characterizes thermal-noise fluctuations
  conditional on one fiducial light cone. It does not include
  cosmic variance, independent light-cone variance, calibration
  errors, or foreground residuals.
}
\\[3.0em]

\hline

\end{tabular}

\endgroup
\end{table*}

From this common parent vector, we construct five statistic
choices: PS alone, $S_1$ alone, $R$ alone, the combined
$S_1+R$ vector denoted WST, and the combined PS+WST vector.
The foreground-wedge calculation remains a coefficient-level
response test and is not included in the Fisher covariance.

Parameter derivatives are estimated using a four-point
linear fit to one-parameter sweeps around the fiducial model.
The same five thermal-noise seeds are used at every
parameter-grid point, reducing noise fluctuations in the
estimated parameter responses. The covariance is measured
directly from 150 transferred and noisy fiducial maps with the
same map geometry and feature extraction used for the
derivatives.

For each statistic choice, the features are standardized
and an independent Ledoit-Wolf covariance is fitted. Fisher
products are evaluated using Cholesky solves rather than explicit
matrix inversion. The covariance is held fixed across the
parameter space. It therefore does not include cosmic variance,
variation among independent signal realizations, or
parameter-dependent covariance. Sample-count, derivative-step,
and regularization tests are presented in
Appendix~\ref{app:validation}.

\subsection{Local Information from PS, WST, and PS+WST}
\label{subsec:fisher_results}

We report the results in fractional physical coordinates,
using $\Delta m_{22}/m_{22,{\rm fid}}$,
$\Delta\zeta/\zeta_{\rm fid}$, and
\begin{equation}
{\color{black}
\frac{\Delta X}{X}
\simeq
\ln 10\,\Delta\ell_\mathrm{X},
\qquad
X=L_\mathrm{X}/{\rm SFR}.
}
\end{equation}
Here $\Delta\ell_{\rm X}$ denotes a displacement from the fiducial logarithmic X-ray parameter, and $\Delta X/X$ is the corresponding fractional displacement of the physical parameter $X=L_{\rm X}/{\rm SFR}$.

Figure~\ref{fig:fisher_primary_errors} compares the
marginalized local errors obtained from the five statistic
choices. Relative to PS alone, the combined PS+WST vector
reduces the errors on
$(m_{22},L_\mathrm{X}/{\rm SFR},\zeta)$
by factors of 1.79, 1.79, and 1.82, respectively. WST alone
improves the corresponding errors by factors of 1.55, 1.54,
and 1.59. The relative improvement is stable across the tested
derivative estimators, varying by less than 0.2\% for
$m_{22}$ and $L_\mathrm{X}/{\rm SFR}$ and by approximately 3.1\% for
$\zeta$. These results show that the WST adds scale-dependent
and redshift-dependent information that is not fully captured
by the selected two-dimensional PS vector.

Table~\ref{tab:absolute_errors} lists the corresponding
local marginalized errors. For the logarithmic X-ray coordinate,
we convert to a fractional physical uncertainty using
$\sigma_X/X=\ln 10\,\sigma_{\ell_\mathrm{X}}$. Several marginalized
errors extend beyond the parameter intervals used to estimate
the numerical derivatives. They should therefore be interpreted
as local Gaussian curvature measures rather than as global
posterior credible intervals or survey constraints.

For the same reason, we do not present Fisher confidence
ellipses. Since the inferred error scales are substantially
broader than the local derivative grid, the resulting ellipses
could be mistaken for globally valid Gaussian posterior regions.
We instead report the marginalized errors and parameter
correlation coefficients directly.

The WST improves the marginalized local constraints by
adding scale-dependent and redshift-dependent responses to the
PS information. At the same time, the principal mass-heating
degeneracy remains strong. The marginalized correlation is
$r(m_{22},\ell_\mathrm{X})\simeq-0.96$
for PS, WST, and PS+WST. Thus, adding the WST narrows the
locally allowed parameter region without fully removing or
strongly rotating the dominant degeneracy direction. The
mass-ionization correlations also change only modestly.

The adopted legacy mass-independent source prescription
does not specify $f_\star$ as an independent numerical parameter.
Instead, $\zeta$ combines the effects of star formation, the
escape fraction of ionizing photons, the stellar ionizing-photon
yield, and recombinations. The present Fisher analysis therefore
does not separately test a degeneracy between $m_{22}$ and an
independently varied star-formation efficiency.

\begin{figure}[htbp]
\centering
\includegraphics[width=1.0\linewidth]
{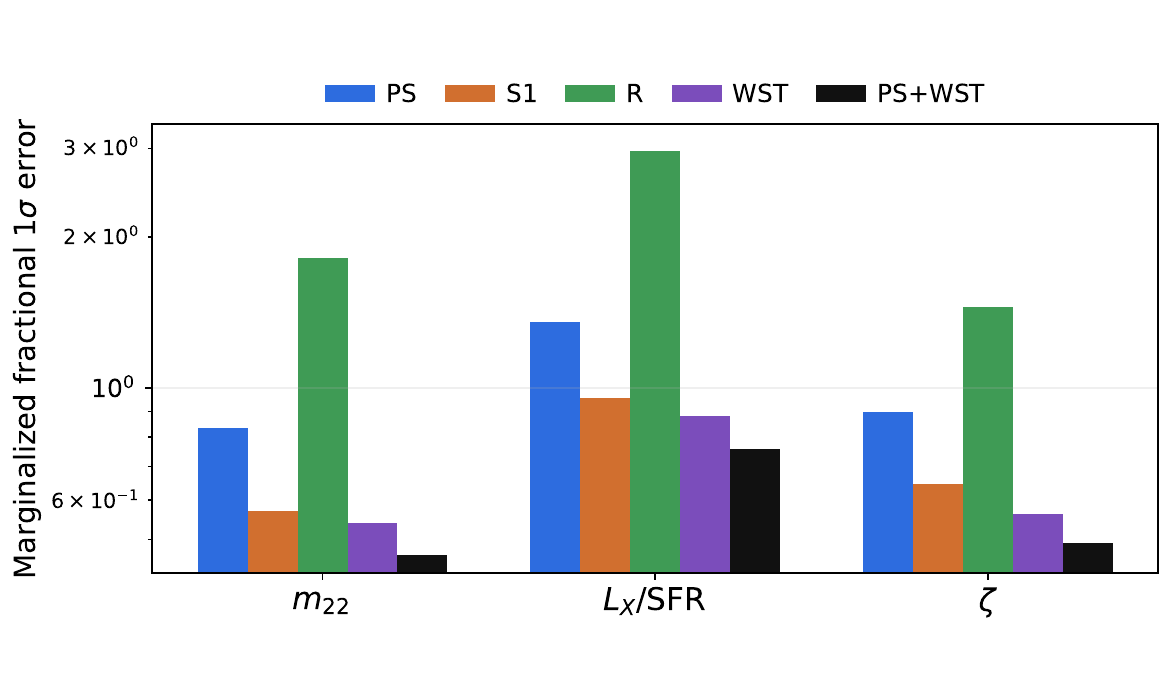}
\caption{Marginalized fractional errors for the primary
17-slab, three-parameter comparison. All statistic choices are
derived from the same map ensemble and use statistic-specific
regularized covariance estimates. The covariance characterizes
thermal-noise fluctuations conditional on one fiducial light cone
and does not include cosmic variance. The plotted values are local
information measures and should not be interpreted as an
end-to-end survey reach.}
\label{fig:fisher_primary_errors}
\end{figure}

\subsection{Redshift Dependence and Interpretation}
\label{subsec:redshift_information}

The primary Fisher range is $8\leq z\leq25$. Within the
adopted atomic-cooling source model, the intervals
$8\leq z<12$ and $12\leq z<20$ contain most of the local
response, while the three slabs at $20\leq z\leq25$ contribute
comparatively little. Adding the three slabs at $z>25$ changes
the PS+WST errors by less than 0.5\%. This test shows that the
highest-redshift slabs do not drive the reported information
gain. It does not demonstrate that molecular-cooling minihalos
or Population III sources are negligible at high redshift.

The choice $z\geq8$ defines the modeling domain of this
study and should not be interpreted as evidence that the WST is
intrinsically more sensitive to density or X-ray-heating
fluctuations than to ionization fluctuations. Across the
analyzed range, the WST responds to the combined morphology
produced by density fluctuations, spatially varying
Ly$\alpha$ coupling, inhomogeneous X-ray heating, and the
development of ionized regions. Because the signal is not
decomposed into these individual components, the information
gain cannot be assigned uniquely to one physical field.

Ionized-bubble morphology is expected to provide an
important source of non-Gaussian structure during the later
stages of reionization. The present analysis does not fully
explore this ionization-dominated regime. We exclude $z<8$
because it lies outside the Cosmic Dawn and early-EoR focus of
the adopted simulation and foreground treatment. A
component-separated analysis extended to lower redshift would
be required to compare the relative WST sensitivity to
heating-driven and ionization-driven non-Gaussianity.

The present calculation is deliberately constructed as a
matched comparison between the WST and the two-dimensional PS
extracted from the same map products. Both statistics use
identical redshift slabs, retained transverse modes,
instrumental transfer, thermal-noise treatment, and parameter
model. This controlled design isolates the complementary
information contributed by the WST relative to the selected
two-dimensional PS baseline.

A fully matched comparison with a three-dimensional or
cylindrical PS would additionally require line-of-sight
correlations, a chromatic visibility-domain response, and the
corresponding covariance construction. The present results
should therefore not be interpreted as establishing that the
two-dimensional WST outperforms a full three-dimensional PS.
Instead, they show that the WST contributes complementary local
morphological information beyond the matched two-dimensional
PS under the adopted setup. The absolute errors remain
conditional on the restricted three-parameter model, one
fiducial light cone, the analytic instrumental transfer model, and a
parameter-independent thermal-noise covariance.

\FloatBarrier
\section{Summary and Discussion}
\label{sec:discussion}

We applied WST to matched 2\,MHz map products from FDM-modified 21\,cm light cones. The first-order block $S_1$ summarizes band amplitudes, while $R=S_2/S_1(j_1)$ captures ordered cross-scale modulation. FDM shifts both summaries indirectly through the HMF and delayed radiative-source history rather than through a direct measurement of the linear cutoff.

Using the same transferred and noisy maps for all summaries, PS+WST reduces the local marginalized errors on $(m_{22},L_\mathrm{X}/{\rm SFR},\zeta)$ by approximately a factor of 1.8 relative to the matched two-dimensional PS. The dominant mass-heating correlation remains near $-0.96$, so the gain reflects complementary response information rather than removal of the principal degeneracy. The absolute errors are order unity and are local Gaussian information measures conditional on one fiducial light cone, not survey constraints.

The analytic instrumental-transfer model, thermal-noise model, and idealized wedge masks are controlled diagnostics rather than an end-to-end SKA-Low pipeline. The wedge test finds a smaller median coefficient response for $R$ than for $S_1$, but does not establish foreground-aware detectability. A full assessment requires independent initial conditions, improved source-model provenance, a physical chromatic visibility model, foreground residuals, and a matched comparison with three-dimensional statistics.

\appendix

\section{Background Equations for Fuzzy Dark Matter}
\label{app:fdm_background}

This appendix summarizes the field-theory background that motivates the FDM cutoff used in the main text. These equations are not solved directly by the semi-numerical 21\,cm pipeline, which uses the modified linear power spectrum and HMF prescription described in Section~\ref{sec:FDM}.

\paragraph{Metric and conventions.}
We use signature $(+,-,-,-)$, $x^\mu=(x^0,\boldsymbol{x})$ with $x^0=ct$,
and comoving $\boldsymbol{x}$:
\begin{equation}
ds^2=\left(1+\frac{2\Phi}{c^2}\right)(dx^0)^2
-a^2(t)\left(1-\frac{2\Phi}{c^2}\right)d\boldsymbol{x}^2.
\label{eq:metric}
\end{equation}
Here $ds^2$ is the spacetime line element, $t$ is cosmic time, $a(t)$ is the scale factor, $\Phi$ is the Newtonian gravitational potential, and $c$ is the speed of light.

\paragraph{Relativistic action.}
For a real scalar $\phi$ with dimensions of inverse length and particle mass $m_{\rm FDM}$, a dimensionally consistent action is
\begin{equation}
S_\phi=\frac{\hbar}{2}\int d^4x\,\sqrt{-g}\,
\left[g^{\mu\nu}\partial_\mu\phi\,\partial_\nu\phi
-\left(\frac{m_{\rm FDM}c}{\hbar}\right)^2\phi^2\right].
\label{eq:action}
\end{equation}
Here $S_\phi$ is the scalar-field action, $g$ is the determinant of the metric $g_{\mu\nu}$, $g^{\mu\nu}$ is the inverse metric, and $\partial_\mu$ denotes a spacetime derivative.
It yields the following equation of motion, known as the Klein-Gordon equation,
\begin{equation}
\left[\Box+\left(\frac{m_{\rm FDM}c}{\hbar}\right)^2\right]\phi=0.
\end{equation}
Here $\Box\equiv g^{\mu\nu}\nabla_\mu\nabla_\nu$ is the covariant d'Alembert operator. The plus sign of the mass term follows from the stated signature.

\paragraph{Nonrelativistic decomposition.}
Removing the rest-mass oscillation, we write
\begin{equation}
\phi=\sqrt{\frac{\hbar}{2m_{\rm FDM}c}}\left[
\psi e^{-im_{\rm FDM}c^2t/\hbar}+\psi^\ast e^{+im_{\rm FDM}c^2t/\hbar}\right],
\label{eq:NRsplit}
\end{equation}
Here $\psi$ is the slowly varying complex nonrelativistic field, $\psi^\ast$ is its complex conjugate, $t$ is cosmic time, $[\psi]=L^{-3/2}$ gives its dimensions, and $\rho=m_{\rm FDM}|\psi|^2$ is the mass density. After averaging the rapidly
oscillating terms, the leading nonrelativistic Lagrangian density is
\begin{equation}
\begin{split}
\mathcal L_{\rm NR}={}&\frac{i\hbar a^3}{2}
(\psi^\ast\dot\psi-\dot\psi^\ast\psi)
-\frac{\hbar^2 a}{2m_{\rm FDM}}|\nabla\psi|^2
-a^3m_{\rm FDM}\Phi|\psi|^2+a^3\bar\rho\Phi
-\frac{a}{8\pi G}|\nabla\Phi|^2.
\end{split}
\label{eq:Lagrangian}
\end{equation}
Here $\mathcal L_{\rm NR}$ is the nonrelativistic Lagrangian density, a dot denotes a derivative with respect to cosmic time, and $\bar\rho$ is the prescribed homogeneous background. Its explicit term
is required for the perturbation potential to be sourced by
$\rho-\bar\rho$, not by the FRW background itself.

\paragraph{Schr\"odinger-Poisson system.}
Varying Eq.~\eqref{eq:Lagrangian} gives
\begin{equation}
i\hbar\left(\partial_t+\frac{3}{2}H\right)\psi
=-\frac{\hbar^2}{2m_{\rm FDM}a^2}\nabla^2\psi+m_{\rm FDM}\Phi\psi,
\label{eq:SP_expanding}
\end{equation}
and
\begin{equation}
\nabla^2\Phi=4\pi Ga^2(\rho-\bar\rho)
=4\pi Ga^2\bar\rho\delta,
\qquad \rho=m_{\rm FDM}|\psi|^2.
\label{eq:Poisson}
\end{equation}
Here $H=\dot a/a$ is the Hubble parameter, $\nabla$ is taken with respect to the comoving spatial coordinate, and $\delta=(\rho-\bar\rho)/\bar\rho$ is the density contrast.
The redefinition $\widetilde\psi=a^{3/2}\psi$ removes the explicit Hubble
term from Eq.~\eqref{eq:SP_expanding}.

\paragraph{Hydrodynamic form.}
Writing $\psi=\sqrt{\rho/m_{\rm FDM}}\,e^{iS/\hbar}$, where $S$ denotes the phase of the nonrelativistic field $\psi$, and defining the peculiar velocity
$\boldsymbol v=\nabla S/(m_{\rm FDM}a)$, Eqs.~\eqref{eq:SP_expanding} and
\eqref{eq:Poisson} become
\begin{align}
\partial_t\rho+3H\rho+\frac1a\nabla\!\cdot(\rho\boldsymbol v)&=0,\\
\partial_t\boldsymbol v+H\boldsymbol v
+\frac1a(\boldsymbol v\!\cdot\!\nabla)\boldsymbol v
&=-\frac1a\nabla\Phi
+\frac{\hbar^2}{2m_{\rm FDM}^2a^3}\nabla\!\left(\frac{\nabla^2\sqrt\rho}{\sqrt\rho}\right).
\end{align}
Equivalently, define the quantum potential by
\begin{equation}
Q=-\frac{\hbar^2}{2m_{\rm FDM}a^2}\frac{\nabla^2\sqrt\rho}{\sqrt\rho}.
\end{equation}
Here $Q$ denotes the quantum potential. The last force is $-(a m_{\rm FDM})^{-1}\nabla Q$. These scale-factor powers agree
with the comoving linear growth equation in Eq.~\eqref{eq:fdm_linear_growth}.

\section{Reference Global Signal and Power Spectrum}
\label{app:reference_signals}

This appendix collects the global-signal and power-spectrum diagnostics of the underlying FDM-modified \texttt{21cmFAST} simulations. These figures are used as reference checks for the physical timing of Ly$\alpha$ coupling, X-ray heating, and reionization. The main text focuses instead on the WST summaries and on the Fisher comparison with the power spectrum.

\begin{figure*}[htbp]
    \centering
    \includegraphics[width=1.0\hsize]{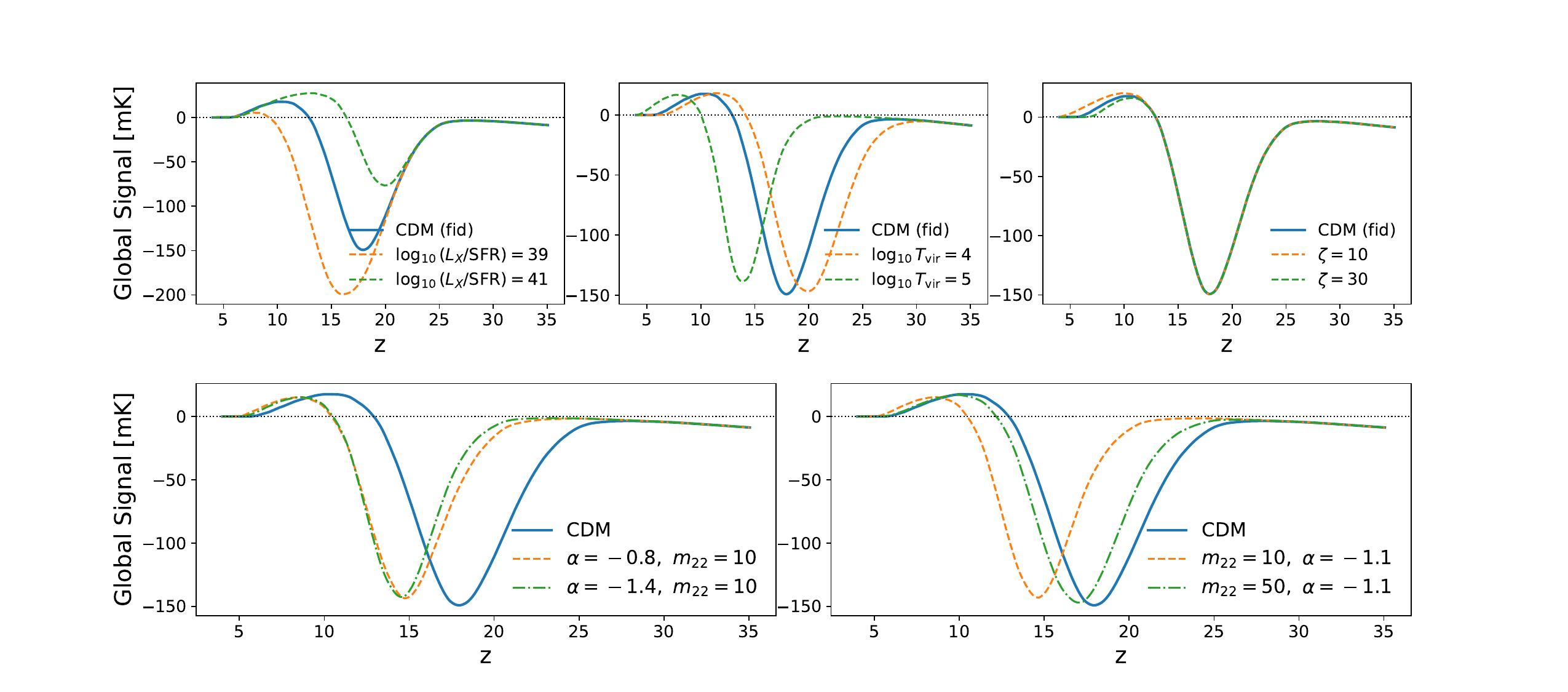}
    \caption{Global 21\,cm signal for representative astrophysical, HMF-shape, and FDM-mass variations. The panels show how changes in $L_\mathrm{X}/{\rm SFR}$, $T_{\rm vir}$, $\zeta$, the HMF-shape parameter $\alpha$, and the FDM mass parameter $m_{22}$ shift the timing and depth of the absorption feature. These trends provide the physical timing reference for the WST evolution discussed in the main text.}
    \label{fig:app_global_signal}
\end{figure*}

\begin{figure*}[htbp]
    \centering
    \includegraphics[width=1.0\hsize]{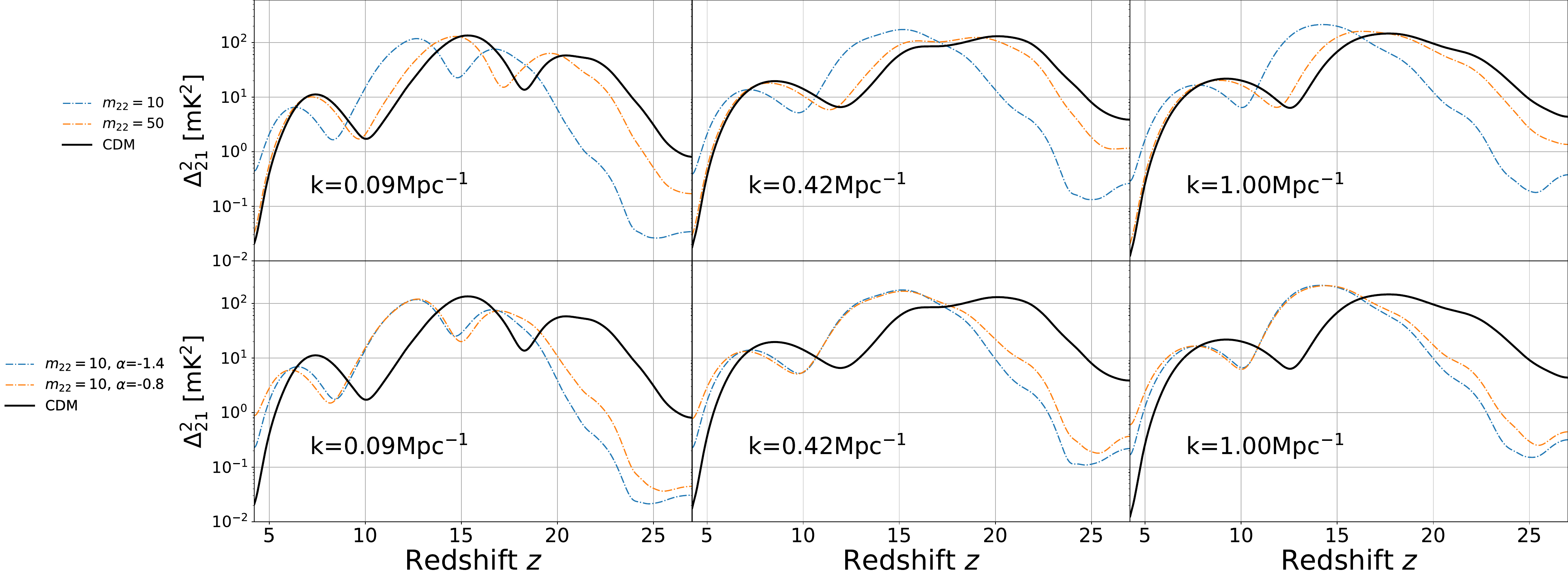}
    \caption{Power spectrum for the FDM mass sweep and the HMF-shape dependence. The top row compares CDM with FDM models with $m_{22}=10$ and $m_{22}=50$, while the bottom row varies the HMF-shape parameter $\alpha$ at fixed $m_{22}=10$. The three columns show representative wavenumbers.}
    \label{fig:app_ps_fdm}
\end{figure*}

\begin{figure*}[htbp]
    \centering
    \includegraphics[width=1.0\hsize]{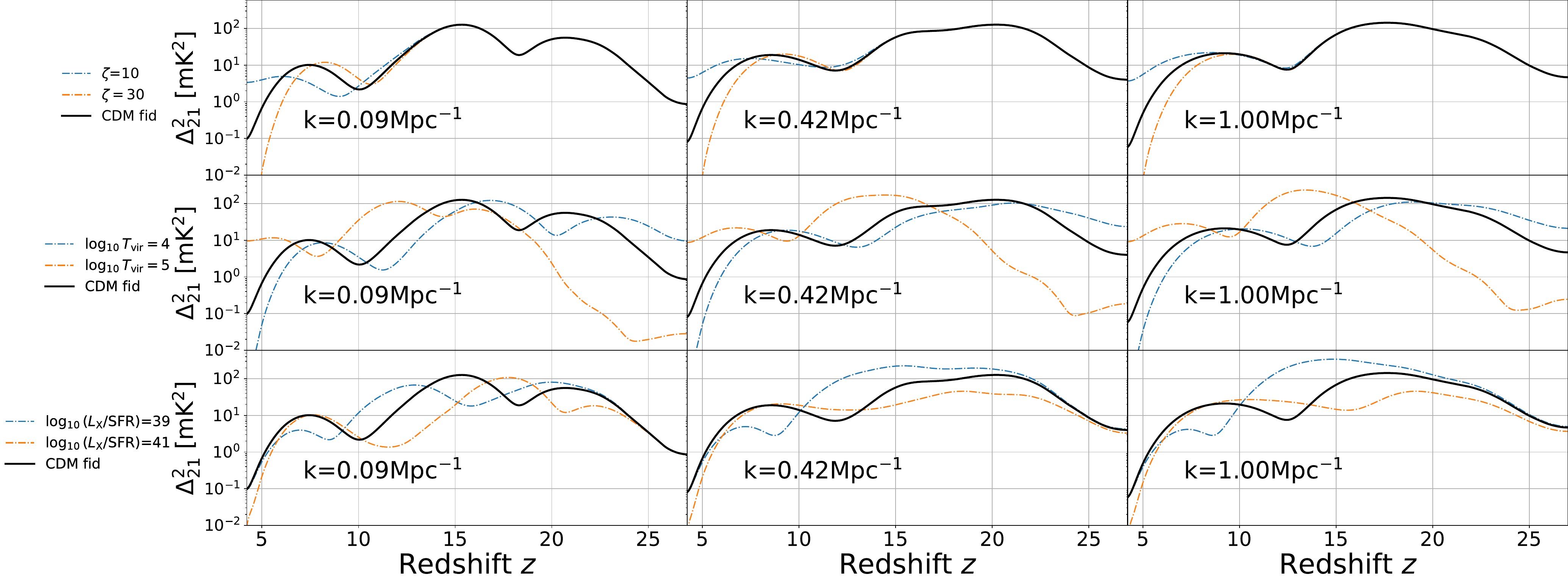}
    \caption{Power spectrum for representative astrophysical parameter variations. The rows vary $\zeta$, $T_{\rm vir}$, and $L_\mathrm{X}/{\rm SFR}$, and the columns show the same representative wavenumbers as in Fig.~\ref{fig:app_ps_fdm}. These trends illustrate how astrophysical timing changes can mimic or compete with the FDM-induced shifts.}
    \label{fig:app_ps_astro}
\end{figure*}
\FloatBarrier

\section{The Morlet Filter Bank}
\label{app:morlet}

The wavelet scattering transform in this work employs a two-dimensional
complex Morlet filter bank as the basic set of analysis functions.
Each wavelet $\psi_{j,\ell}(\mathbf{x})$ is obtained from a mother Morlet function
$\psi(\mathbf{x})$ by scaling and rotation:
\begin{equation}
\psi_{j,\ell}(\mathbf{x})
= 2^{-2j}\,
  \psi\!\left(R_{\theta_\ell}^{-1}\frac{\mathbf{x}}{2^j}\right),
\qquad
\theta_\ell = \frac{\pi \ell}{L},
\label{eq:morlet_scale_rotate}
\end{equation}
where $j=0,\ldots,J-1$ denotes the wavelet scale index and
$\ell=0,\ldots,L-1$ the orientation index.
The operator $R_{\theta_\ell}$ rotates the coordinate system by angle
$\theta_\ell$, allowing the filter bank to detect anisotropic or filamentary
features in multiple directions.
The set $\{\psi_{j,\ell}\}_{j,\ell}$ thus forms the
\textit{filter bank} used in the scattering transform.

For a continuous, isotropic illustration define the normalized Gaussian
\begin{equation}
g_\sigma(\boldsymbol x)=\frac{1}{2\pi\sigma^2}
\exp\left(-\frac{|\boldsymbol x|^2}{2\sigma^2}\right),
\qquad
\beta=\exp\left(-\frac{\sigma^2|\boldsymbol q_0|^2}{2}\right).
\end{equation}
A zero-mean Morlet wavelet and its Fourier transform are then
\begin{align}
\psi(\boldsymbol x)&={\cal N}g_\sigma(\boldsymbol x)
\left[e^{i\boldsymbol q_0\cdot\boldsymbol x}-\beta\right],
\label{eq:morlet_realspace}\\
\widehat\psi(\boldsymbol q)&={\cal N}\left[
e^{-\sigma^2|\boldsymbol q-\boldsymbol q_0|^2/2}
-\beta e^{-\sigma^2|\boldsymbol q|^2/2}\right].
\label{eq:morlet_fourier}
\end{align}
Here $\sigma$ is the Gaussian-envelope width, $\boldsymbol q_0$ is the central wavevector of the Morlet wavelet, $\boldsymbol q$ is the Fourier-space wavevector, ${\cal N}$ is a normalization constant, and $\beta$ is the subtraction coefficient that enforces the zero-mean condition $\widehat\psi(\boldsymbol 0)=0$.
The correction is multiplied by the same Gaussian envelope; subtracting a
spatial constant would be nonintegrable on the plane. Equations
\eqref{eq:morlet_realspace} and \eqref{eq:morlet_fourier} are explanatory
continuous forms. The analysis uses the periodized, anisotropic discrete
filters constructed and internally normalized by Kymatio version
\texttt{0.4.0.dev0} \citep{2018arXiv181211214A}. Here $J$ controls the maximum
dyadic scale, $L$ is the number of orientations, and \texttt{max\_order}
controls the cascade depth; no parameter $Q$ is used as the scattering order.

We measured the physical passbands directly from the padded $256^2$ Kymatio
filters using 1.5 comoving Mpc pixels. For the $J=5$, $L=4$ Fisher bank, the
power-weighted centroids for code indices $j=0,1,2,3,4$ are
$(1.675,0.879,0.440,0.220,0.110)\,{\rm Mpc}^{-1}$. The corresponding
10th to 90th percentile power intervals are $(1.084,2.192)$,
$(0.542,1.227)$, $(0.270,0.613)$, $(0.135,0.306)$, and
$(0.067,0.151)\,{\rm Mpc}^{-1}$. These overlapping finite bands, also shown
in Figure~\ref{fig:transfer}, replace the unverified statement that $J=6$
reaches the box fundamental mode. The $J=6$ descriptive bank adds a $j=5$
band with centroid $0.0549\,{\rm Mpc}^{-1}$ and 10th to 90th percentile range
from $0.0355$ to $0.0763\,{\rm Mpc}^{-1}$; even this does not reach the
$2\pi/300=0.0209\,{\rm Mpc}^{-1}$ transverse fundamental.

The scaling and rotation in Eq.~(\ref{eq:morlet_scale_rotate}) provide the
multiscale and multiorientation coverage used in the analysis. The measured
physical passbands, rather than illustrative one-dimensional wavelet profiles,
are the quantities relevant to the transfer comparison in
Figure~\ref{fig:transfer}.

\FloatBarrier

\section{Dependence on the HMF-shape Parameter}
\label{app:alpha_robustness}

This appendix collects the tests in which we vary the HMF-shape parameter $\alpha$ at fixed $m_{22}=10$. The parameter $\alpha$ is part of the phenomenological FDM HMF prescription and controls the shape of the low-mass suppression. It is therefore a meaningful modeling parameter of the FDM halo sector, although not a fundamental particle-physics parameter like $m_{22}$. In the present Fisher analysis we keep $\alpha$ fixed and use these tests to show how the WST summaries respond to this nuisance direction in the adopted HMF model.

\begin{figure*}[htbp]
    \centering
    \begin{minipage}[t]{0.49\textwidth}
      \centering
      \includegraphics[width=\linewidth]{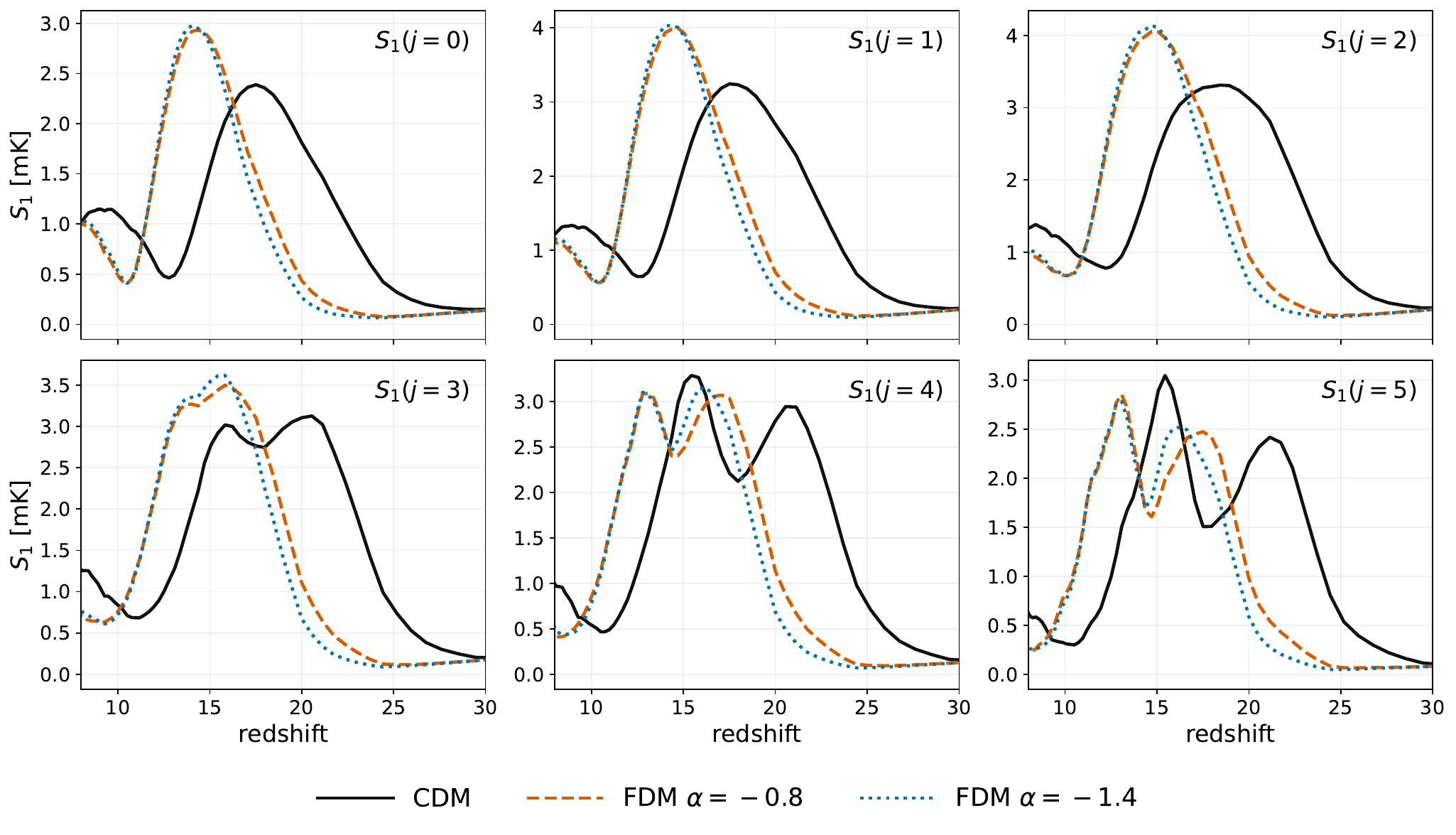}\\[-0.4ex]
      {\small\bfseries (a)}
    \end{minipage}\hfill
    \begin{minipage}[t]{0.49\textwidth}
      \centering
      \includegraphics[width=\linewidth]{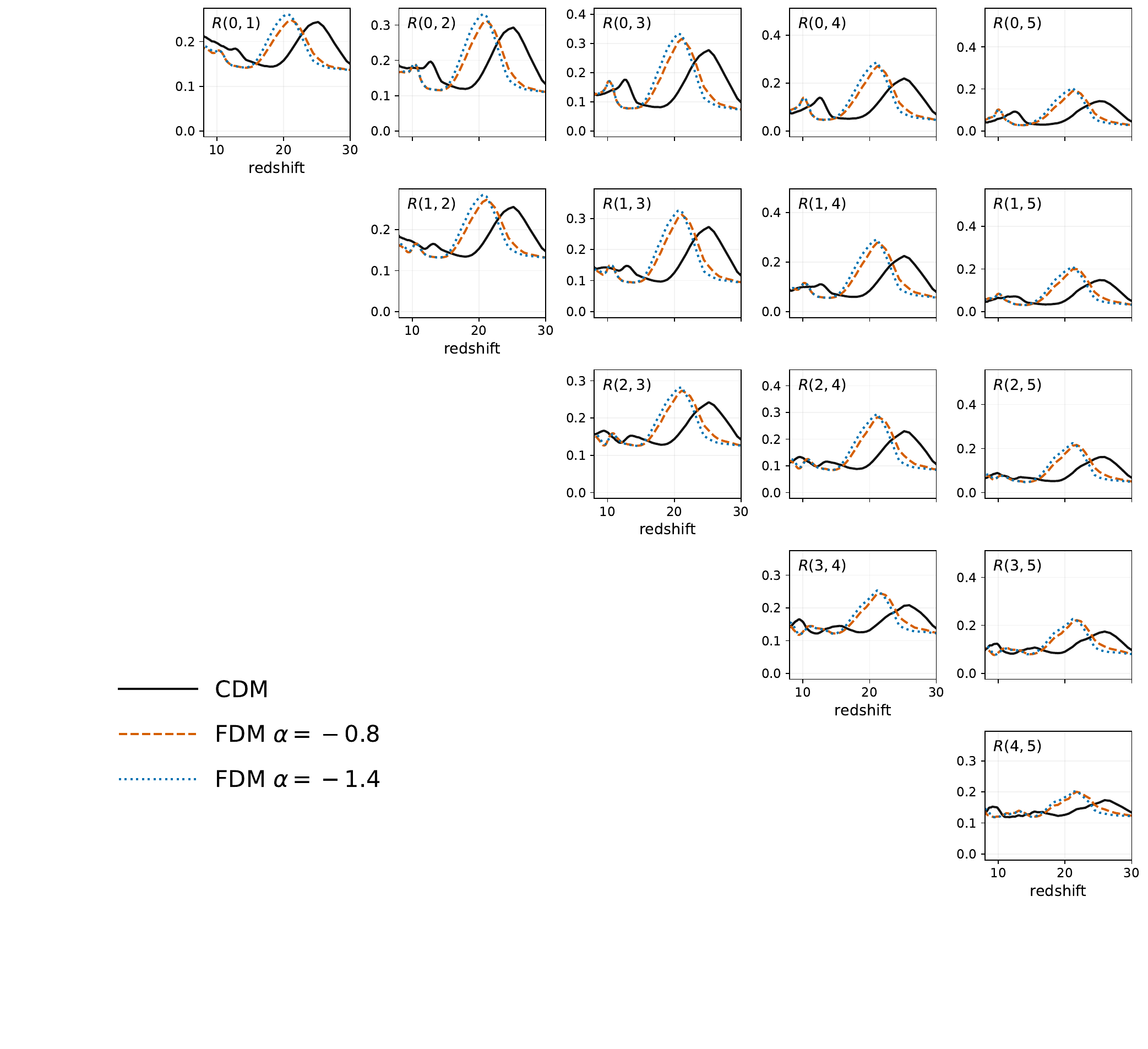}\\[-0.4ex]
      {\small\bfseries (b)}
    \end{minipage}
    \caption{Dependence on the phenomenological HMF-shape parameter $\alpha$ at fixed $m_{22}=10$. Panel (a) shows $S_1$ and panel (b) shows the ordered ratio $R=S_2/S_1(j_1)$. Both use metadata-based coefficient extraction with $S_0$ excluded. These are descriptive sensitivity tests; $\alpha$ is fixed in the primary Fisher analysis.}
    \label{fig:wst_1st_alpha_app}
    \label{fig:wst_2nd_alpha_app}
\end{figure*}

\FloatBarrier

\section{Alternative Second-Order Normalization}
\label{app:normalization}

We compare the primary ordered normalization
\begin{equation}
R(j_1,j_2)=\frac{S_2(j_1,j_2)}{S_1(j_1)}
\end{equation}
with the symmetric normalization
\begin{equation}
\widetilde R(j_1,j_2)=\frac{S_2(j_1,j_2)}{\sqrt{S_1(j_1)S_1(j_2)}}
=R(j_1,j_2)\sqrt{\frac{S_1(j_1)}{S_1(j_2)}}.
\end{equation}
The scattering path remains directed in both cases; $\widetilde R$ changes only the amplitude normalization and is not a conventional correlation coefficient. Table~\ref{tab:normalization_comparison} shows that $R$ has the smaller noise coefficient of variation, while the wedge response depends on the mask. The symmetric normalization gives smaller local Fisher errors in this simulation set, but neither definition is uniformly preferable across the diagnostics. We retain $R$ as the primary statistic because it follows the parent-response normalization of the ordered cascade and treat $\widetilde R$ as a sensitivity test.

\begin{table*}
\centering
\caption{Sensitivity to the second-order normalization.
The wedge-response values are ordered as optimistic,
horizon, and buffered horizon. The fractional-error
and improvement tuples are ordered as
$(m_{22},L_\mathrm{X}/{\rm SFR},\zeta)$. The fractional errors are the
marginalized PS+WST errors obtained when the corresponding
normalization is used in the combined data vector.}
\label{tab:normalization_comparison}

\begingroup
\footnotesize
\setlength{\tabcolsep}{7pt}
\renewcommand{\arraystretch}{1.25}

\begin{tabular}{@{}lcccc@{}}
\toprule
Statistic
&
Noise CV
&
Wedge response
&
Fractional errors
&
$\sigma_{\rm PS}/\sigma_{\rm PS+WST}$
\\
\midrule
$R$
&
0.0185
&
$(0.174,0.261,0.247)$
&
$(0.466,0.757,0.493)$
&
$(1.79,1.79,1.82)$
\\
$\widetilde R$
&
0.0336
&
$(0.222,0.239,0.220)$
&
$(0.375,0.637,0.419)$
&
$(2.23,2.12,2.14)$
\\
\bottomrule
\end{tabular}

\endgroup
\end{table*}

\section{Covariance and Derivative Validation}
\label{app:validation}

The primary PS+WST vector has 340 components and is estimated from 150 direct noisy-map realizations, so an unregularized sample covariance is rank deficient. Each data-vector choice is standardized and assigned an independent Ledoit-Wolf covariance estimate; the fitted shrinkage coefficients range from 0.47 to 0.97, and the regularized standardized matrices have condition numbers below 5.8. Fisher products use Cholesky solves.

The regularization is stable under the available checks. Reducing the covariance ensemble from 150 to 120 samples changes the combined errors by 3.2\%, 3.9\%, and 0.1\% for $(m_{22},L_\mathrm{X}/{\rm SFR},\zeta)$, while removing cross-statistic covariance blocks changes them by about 2\%. Derivatives estimated from the inner $\pm1\%$ pair, outer $\pm2\%$ pair, and four-point fit change every marginalized error by less than 3.3\%. Individual $R$ components are less stable, especially for $\zeta$, so we interpret only the full covariance-weighted data vector.

Table~\ref{tab:absolute_errors} lists the local marginalized
Fisher errors. For the logarithmic X-ray coordinate, we convert
to a fractional physical uncertainty using
$\sigma_X/X=\ln 10\,\sigma_{\ell_\mathrm{X}}$. Because several marginalized
errors extend beyond the parameter intervals used to estimate the
numerical derivatives, these values should be interpreted as local
Gaussian curvature measures rather than as global posterior
credible intervals. The covariance used here characterizes thermal-noise
fluctuations conditional on one fiducial light cone and is held fixed
across parameter space. It therefore does not include cosmic variance,
variation among independent light-cone realizations, or
parameter-dependent covariance. Quantifying these additional
contributions would require independent signal realizations and
covariance estimates across the parameter space. We therefore use
the absolute errors cautiously and emphasize the relative information
comparison among the matched PS, WST, and PS+WST data vectors.

\begin{table*}
\centering
\caption{
Marginalized fractional $1\sigma$ errors for the primary
three-parameter local Fisher comparison. WST denotes
$S_1+R$, while PS+WST denotes the combined power-spectrum
and WST data vector.}
\label{tab:absolute_errors}
\footnotesize
\setlength{\tabcolsep}{9pt}
\renewcommand{\arraystretch}{1.25}

\begin{tabular}{@{}lccccc@{}}
\toprule
Parameter
&
PS
&
$S_1$
&
$R$
&
WST
&
PS+WST
\\
\midrule
$m_{22}$
&
0.835
&
0.570
&
1.81
&
0.540
&
0.466
\\
$L_\mathrm{X}/{\rm SFR}$
&
1.35
&
0.955
&
2.96
&
0.880
&
0.757
\\
$\zeta$
&
0.897
&
0.645
&
1.45
&
0.564
&
0.493
\\
\bottomrule
\end{tabular}

\end{table*}

\FloatBarrier

\begin{acknowledgments}
BL is supported by the Guangxi Key R\&D Program (Guangxi Funeng Action Plan, Guike~FN2504240040), the Guangxi Natural Science Foundation (grant No.~2026GXNSFHA00640301), the Guangxi Science and Technology Innovation Platform Program (Leitai Action Plan, grant No.~Guike~LT2600640026) and the National Natural Science Foundation of China (grant Nos.~12203012 and 12494575). HS is supported by the National SKA Program of China (No.~2020SKA0110401), NSFC (Grant No.~12103044), and Yunnan Provincial Key Laboratory of Survey Science with project No.~202449CE340002. Additional support was provided by the Guangxi Talent Program (``Highland of Innovation Talents'').
\end{acknowledgments}

\bibliographystyle{aasjournalv7}
\bibliography{reference}
\label{lastpage}
\end{document}